\documentclass[preprint2]{aastex}

\shorttitle{Visual Binaries in the Orion Nebula Cluster}
\shortauthors{Reipurth et al.}

\begin{document}


\title{Visual Binaries in the Orion Nebula Cluster}


\author{Bo Reipurth\altaffilmark{1}, Marcelo M. Guimar\~aes\altaffilmark{1,2}, Michael S. Connelley\altaffilmark{1,3}, and
John Bally\altaffilmark{4}}


\altaffiltext{1}{Institute for Astronomy, University of Hawaii, 640
  N. Aohoku Place, Hilo, HI 96720 (reipurth@ifa.hawaii.edu)}
\altaffiltext{2}{Departamento de Fisica, UFMG, Caixa Postal 702,
  30.123-970, Belo Horizonte, MG, Brazil } 
\altaffiltext{3}{NASA Ames Research Center, Moffett Field, CA 94035}
\altaffiltext{4}{CASA, University of Colorado, 389 UCB, Boulder,
  Colorado 80309-0389 }


\begin{abstract}

We have carried out a major survey for visual binaries towards the
Orion Nebula Cluster using HST images obtained with an H$\alpha$
filter.  Among 781 likely ONC members more than 60$''$ from
$\theta^1$~Ori~C, we find 78 multiple systems (75 binaries and 3
triples), of which 55 are new discoveries, in the range from
0$\farcs$1 to 1$\farcs$5.  About 9 binaries are likely line-of-sight
associations. We find a binary fraction of 8.8\%$\pm$1.1\% within the
limited separation range from 67.5 to 675 AU.  The field binary
fraction in the same range is a factor 1.5 higher. Within the range
150 AU to 675 AU we find that T~Tauri associations have a factor 2.2
more binaries than the ONC.  The binary separation distribution
function of the ONC shows unusual structure, with a sudden steep
decrease in the number of binaries as the separation increases beyond
0$\farcs$5, corresponding to 225~AU. We have measured the ratio of
binaries wider than 0$\farcs$5 to binaries closer than 0$\farcs$5 as a
function of distance from the Trapezium, and find that this ratio is
significantly depressed in the inner region of the ONC.  The deficit
of wide binaries in the central part of the cluster is likely due to
dissolution or orbital change during their passage through the
potential well of the inner cluster region.  Many of the companions
are likely to be brown dwarfs.

\end{abstract}





\section{INTRODUCTION}

Since the first detection of pre-main sequence binaries by Joy \& van
Biesbroeck (1944) and Herbig (1962), major advances have been gained
in our understanding of the formation and early evolution of
binaries. One of the key results is that binaries are about twice as
common in T Tauri associations as among field stars (e.g., Reipurth \&
Zinnecker 1993, Simon et al. 1995, Duch\^ene 1999, Ratzka et
al. 2005). Studies of embedded sources suggest that multiple systems
may possibly be more prevalent among Class~I sources than among the
more evolved T Tauri stars (e.g., Duch\^ene et al. 2004).  Dynamical
evolution among higher-order multiples is likely to be important
during the embedded phase, leading to the occasional ejection of
lower-mass members and the formation of powerful jets (Reipurth 2000).

The Orion Nebula Cluster (ONC) is the nearest (d$\sim$450~pc) star
forming cluster, with an age of less than 1 Myr, and it has been
extensively studied at multiple wavelengths (e.g., Herbig \& Terndrup
1986, Hillenbrand 1997, Getman et al. 2005 and references therein; for
earlier work see the review by Herbig 1982). It is well known that the
massive stars in the Trapezium have a very high binary frequency (e.g., 
Preibisch et al. 1999). Binarity of the low-mass population of the
central part of the ONC has been studied extensively, and it has been
suggested that the binary frequency is lower than for the dispersed
low-mass young population in associations and comparable to the binary
frequency in the field (Prosser et al. 1994, Padgett et al. 1997, Petr
et al. 1998, Simon et al. 1999, K\"ohler et al. 2006).

In this paper, we report the results of a major binary survey of an
extensive region centered on the ONC, based on a large H$\alpha$
imaging survey using the Hubble Space Telescope. We have detected 78
multiple systems, of which 55 are new discoveries, in an area of 412
arcmin$^2$ that excludes a circular area with radius of 60$''$
centered on $\theta^1$~Ori~C.  We discuss the membership of these
binaries in the ONC, analyze their binary properties, and discuss the
binary separation distribution function and binary formation in the
ONC.

\section{OBSERVATIONS}





The present survey is based on our recent study of the Orion Nebula
using the Wide Field Camera of the Advanced Camera for Surveys onboard
HST. During program GO-9825, we observed 26 ACS fields with the F658N
(H$\alpha$+[NII]) filter and with an exposure time of 500 sec per
pointing.  In total our H$\alpha$ survey covers an area of 415
arcmin$^2$ of the central Orion Nebula. Due to the high stellar
density near the Trapezium we have excluded a circular area with
radius of 60$''$ centered on $\theta^1$~Ori~C, so the area
investigated is 412 arcmin$^2$.  The exquisite resolution of the HST
combined with the 0$\farcs$05 pixel-size of the ACS offers a unique
opportunity to detect close binaries. We examined all stellar sources
in the images by eye at a range of contrast levels, allowing us to
pick up faint as well as bright companions. Thanks to the very stable
point-spread function across the images, this could be done in a
rather homogeneous fashion. We discuss the completeness of this
procedure in Sect.~3.3.  We used reduced images that were combined
through MULTIDRIZZLE.  The faintest star we measured had a magnitude
of 21.5 in the F658 filter. Pixel coordinates for both primaries and
secondaries were determined with 2-dimensional Gaussian functions, and
separations and position angles calculated from these coordinates.
Relative fluxes were determined using aperture photometry with
corrections for the sky background.  For further details of the
observations, see Bally et al. (2006).

\section{SELECTION OF SAMPLE}

\subsection{Survey Area} 

The region of the ONC we have studied is shown in Figure~1, which
outlines the mosaic of 26 ACS fields. One field had to be slightly
turned to acquire a guidestar. We examined all stars (1051) in these
fields with the exception of stars within an exclusion zone of 60$''$
around $\theta^1$~Ori~C. 

\subsection{Separation Limits} 

All previous HST studies of binarity in the ONC were done in the
region around the Trapezium, where line-of-sight pairs due to the high
stellar density is a major issue, and thus focused mostly on
sub-arcsec binaries. In contrast, our study covers a much larger area
including the outskirts of the ONC. We have also excluded the inner
region with radius of 60$''$, so contamination is much less of an
issue. Based on our observed star density, we can calculate the
fraction of our binaries that are not physical pairs for different
separation limits. From this we have chosen an upper separation limit
of 1$\farcs$5, which implies an 11\% contamination, as described
below.

\subsection{Incompleteness}  

We have tested our completeness of companions by blindly ``observing''
real stars with artificially added companions (scaled from real stars
in the images) at random separations, position angles, and flux
ratios. We find that our detections are essentially complete down to
separations of 0$\farcs$1, a result made possible by the very sharp
point spread function of the HST/ACS system.  Figure~2 shows two
panels with plots of $\Delta$m vs. separation in arcseconds.  The left
panel shows the actual observations of the binaries we detected, while
the right panel shows our artificial binaries. Filled circles indicate
companions we were able to identify, and empty circles indicate those
we failed to see. Our detection limit is essentially complete to
0$\farcs$1 except for the largest brightness differences, but very few
binaries are found with large flux differences (see Sect. 4.5). To be
certain that we are complete, we considered binary companions in the
more limited separation range from 0$\farcs$15 to 1$\farcs$5.
However, a much more severe problem with completeness is due to the
strong and highly structured emission from the surrounding HII region,
a problem that is compounded by our use of images taken through an
H$\alpha$ filter. In some areas, especially near the Trapezium, our
ability to detect faint stars is diminished, an effect that is
unquantifiable but should be kept in mind.

Most binaries that have been found prior to this study were identified
using infrared techniques. Whereas all binaries we have detected have
been checked against lists of previously discovered binaries, we have
not made any systematic effort to check if all these earlier binaries
are also detectable in the optical. From random checks it appears that
many of these infrared-detected binaries are not visible in our
optical images.

\subsection{Membership of the ONC}  

The main source for membership information in the ONC are astrometric
studies, principally the study by Jones \& Walker (1988), which covers
the inner 20$'$$\times$30$'$ of the ONC and thus includes our area of
study. It is based on photographic I-band plates, which favors red and
reddened stars.  Of the 1051 stars in our images, 655 stars are listed
in the Jones \& Walker catalog, and of these 596 have membership
probabilities higher than 93\%. The large majority of our binaries are
not resolved in the JW survey, and thus the proper motions can be
affected by relative brightness variations of the components that can
shift the photocenter.  We have seen such variability in a binary
lying in the overlap region between two fields, and thus observed
twice.  We have also found stars with clear pre-main sequence
characteristics that have membership probability of 0\%, indicating
that astrometric information can be used to {\em support} membership,
but should not be used to {\em exclude} membership. We are therefore
complementing the proper motion membership analysis with other sources,
such as X-ray emission, H$\alpha$ emission, and variability.  Getman
et al. (2005) has determined that 1373 stars from the 1616 sources in
the COUP survey, which largely overlaps with our field (see Figure~1),
are Orion members.  Among our stars, we find 658 which are detected by
COUP and also are classified as ONC members.  Finally, we use the
presence of H$\alpha$ emission (partly from the literature and from
our own unpublished deep survey), as well as irregular variability as
listed in the General Catalogue of Variable Stars. A total of 99 stars
were found to have H$\alpha$ emission.  Note that we do not use
near-infrared excess as a membership criterion, since our binaries by
default are likely to have infrared excesses due to their mostly
lower-mass companions. In total, we find 781 stars that have at least
one (and commonly several) of these membership characteristics.

\subsection{Binaries and Line-of-Sight Pairs}

Among the 781 known ONC members detected in our HST images, we found
72 binaries and 3 triples in the range from 0$\farcs$15 to
1$\farcs$5. These are listed in Table~1, where each triple is listed
as two binaries, in the manner of counting of Kuiper (1942). The first
and second columns list the Jones \& Walker ID and the probability of
the star to be a member of the ONC based on its proper motion,
respectively.  The third column lists the Parenago number, the fourth
column lists the name in the General Catalog of Variable Stars, the
fifth column lists the COUP number, and the sixth column lists the
number given by Hillenbrand (1997). The right ascension and
declination for equinox 2000 are listed in the seventh and eighth
columns. The ninth column lists the position angle of the system and
the tenth column lists its separation in arcsec.  The apparent
magnitude in the I band (from Hillenbrand 1997) is listed in the
eleventh column and the difference in magnitudes, in the H$\alpha$
filter, between the primary and secondary is listed in the twelfth
column. The position angle towards and separation (in arcsec) to
$\theta^1~$Ori~C are listed in columns 13 and 14, respectively.  Then
follows the spectral type as listed by SIMBAD.  The sixteenth column
lists the character of membership for each binary (P: proper motion;
X: COUP ONC source; H: H$\alpha$ emission; V: irregular variable). The
last column lists the discovery paper if a binary was known prior to
this work.

Additionally, we have found 3 binaries among the ONC members outside
the 60$''$ exclusion zone with separations between 0$\farcs$11 and
0$\farcs$15. They are listed in Table~2.  We have furthermore examined
the optically visible ONC members within the exclusion zone, and have
found another 3 binaries with separations less than 0$\farcs$4, an
(arbitrarily chosen) upper limit we imposed due to the high stellar
density in that region. Two of these are new discoveries. Many more
binaries have been found by earlier studies of the
central region (e.g., Prosser et al. 1994, Padgett et al. 1997, Petr
et al. 1998, Simon et al. 1999, K\"ohler et al. 2006), but most are
only detectable in the infrared, since most cluster members are hidden
by extinction; these three binaries are the only we could detect in
our images in the range 0$\farcs$1 - 0$\farcs$4.  We do not include
the abovementioned binaries in our statistical analysis, but list them
in Table~2. Finally, we have searched the 270 stars outside the
exclusion zone for which we have no evidence for membership, and have
found another 7 binaries. Some of these may be foreground or
background stars, but others may turn out to be young stars, and we
also list them in Table~2.  The columns in Table~2 are the same as in
Table~1, except we have added an additional column, which
characterizes the binaries.

All of our binaries with known spectral types are late-type stars,
presumably classical and weak-line T Tauri stars, with the exception
of Parenago~2149 (JW~945), which appears to be a Herbig Ae/Be star.

In Figure~3a,b we show figures of the 78 multiple systems we have
identified among ONC members outside the exclusion zone, as well as
the three close binaries ($<$0$\farcs$4) found inside the exclusion
zone. Each stamp is 2$''$ on a side. Some of the binaries are
particularly interesting for a variety of reasons, e.g. V1118 Ori is a
famous EXor, other binaries or companions are likely substellar
objects, and some are associated with jets or proplyds. In the
Appendix, we provide more details about these cases.

Could some of the wider binaries in the ONC be due to contamination by
line-of-sight pairs?  The surface density of stars in the ONC is a
steeply declining function of distance from the center of the cluster,
with only relatively minor fluctuations due to subclustering (either
real or caused by extinction variations, Figure~4).  We have
determined the stellar density $\Sigma$ in an area with radius 30$''$
around each of the 781 ONC members, and have determined the
probability $P$ of finding an unrelated star within a distance
$\theta$ from each primary using the expression $P = 1 -
e^{-\pi\theta^2\Sigma}$ (Correia et al. 2006), where we have set
$\theta$ to 1$\farcs$5. The probability $P$ is principally a function
of the distance from $\theta^1$~Ori~C, with smaller variations due to
local inhomogeneities in the cluster density.  Figure~5 shows the
distribution of probability of a line-of-sight association for each of
the 781 ONC members as a function of distance from
$\theta^1$~Ori~C. The figure mostly follows the radial stellar density
curve in Figure~4, although the effect of local small-scale groupings
is visible.

The sum of contamination probabilities for the 781 stars is 911\%.  In
other words, among the binaries detected we are likely to have roughly
9 line-of-sight doubles. In the following we correct for these false
binaries.

\section{RESULTS}

\subsection{Binary Fraction in ONC} 

The detection of 72 binaries and 3 triples with separations in the
range 0$\farcs$15 - 1$\farcs$5 among the 781 ONC members must be
corrected for the estimated 9 binaries that are likely to be due to
line-of-sight pairing. Counting the three triples as six binaries, we
then have 69 physical binaries, which implies a 8.8\%$\pm$1.1\% binary
fraction $bf$ in the interval from 67.5 to 675~AU, or a multiplicity
frequency (where the 3 triples are counted as 1 system each) of
8.5\%$\pm$1.1\% (here and in the following the error estimates are
1$\sigma$ and are derived using Poisson statistics).  Petr et
al. (1998) found 4 binaries in the separation range 0$\farcs$14 to
0$\farcs$5 in the inner 40$''$$\times$40$''$ of the Trapezium,
corresponding to 5.9\%$\pm$4.0\%.  In the same limited range of
separations we find 50 binaries, corresponding to 6.4\%$\pm$0.9\%. The
numbers are consistent within their errors.

Reipurth \& Zinnecker (1993) observed nearby T Tauri associations and
found 38 binaries (of a sample of 238 stars) in the range 150-1800
AU. The range we study statistically is 67.5-675 AU. We have derived
the number of binaries in the {\em common} range 150-675 AU, and find
11.8\%$\pm$2.2\% for the associations and 5.3\%$\pm$0.8\% in the ONC
(for simplicity we here count a triple as two binaries). Both surveys
represent {\em observed} binary fractions, but the Reipurth \&
Zinnecker survey has negligible incompleteness correction in the
chosen range. For our present survey, we have 9 line-of-sight pairs
among our 78 binaries, and when we analyze their statistical
distribution for different separations, we find that all 9 are likely
to be found in the range 0$\farcs$33 to 1$\farcs$5. Of the 50 binaries
in the range 150-675 AU we thus should count only 41 as physical
binaries, leading to the abovementioned binary percentage of 5.3\%.
We thus find that the binary fraction in associations
is higher by a factor of 2.2 compared to the ONC, in qualitative
agreement with the study of Petr et al. (1998), who found a difference
of almost a factor of 3 based on small-number statistics.

\subsection{Separation Distribution Function}  

In Figure~6 we show the separation distribution of binaries
towards the ONC on an angular scale and in bins 0$\farcs$1 wide.  The
first bin from 0$\farcs$1 to 0$\farcs$2 is incomplete. It is striking that
there is clear evidence for structure in the separation distribution
function, with a sudden decrease in the number of binaries as the
separations increase beyond 0$\farcs$5, corresponding to 225~AU.  At
larger separations, the distribution is quite flat.  We return to the
physical interpretation of this ``wall'' in subsequent sections.
An early precursor to this result may have been seen by Simon (1997),
who studied the two-point correlation function in the ONC based on the
results of Prosser et al. (1994) and noted a transition from binary
companions to the large-scale cluster at projected separations around
400~AU.

We have made the same plot on a logarithmic scale in Figure~7. The
data are essentially complete over the whole separation range
displayed. We have corrected the distribution for the 9 line-of-sight
pairs by calculating the probability of finding another star within a
circle with radius corresponding to the separations of each binary,
and then adding these probabilities up for each bin. While such a
procedure is meaningless for an individual object, it can be used to
distribute the 9 false binaries across the 5 bins. Binaries with the
largest separations are most likely to be line-of-sight pairs, and the
last bin has a large correction of 6 stars, while three of the four
remaining bins each is corrected for one star. Error bars are given
for each corrected column.

Figure~7 additionally shows the distribution of field stars in the
same interval from Duquennoy \& Mayor (1991). The dot-dash curve
represents the Gaussian that Duquennoy \& Mayor fitted to their entire
data set, while the two dashed crosses represent their actual data
points within our range. It is evident that the Gaussian is a poor fit
to these data points in this specific separation interval. The
Duquennoy \& Mayor data are given as a function of period, and we have
converted these to separations by assuming a mean total binary mass of
1.2 M$_\odot$, appropriate for their sample of F7 to G9 binaries.  We
have also assumed, as is commonly done, that the projected separation
represents the semimajor axis, although statistically there is a small
difference (Kuiper 1935, Couteau 1960).

As noted above, we find a binary frequency (after correcting for the 9
line-of-sight pairs and treating the three triples as six binaries) of
8.8\%$\pm$1.1\% in the interval from 67.5 AU to 675 AU. If we
calculate the binary frequency from the Duquennoy \& Mayor data in the
same range, we find a binary frequency of 13.7\% using a simple
trapezoidal approximation to their log-normal curve, and 12.4\% using
linear interpolation of their data points. It follows that, in the
specific range from 67.5 AU to 675 AU, there is approximately 1.5
times more binaries in the field than in the ONC. We discuss this
result further below.

\subsection{Wide vs Close Binaries as Function of Distance from the Trapezium}

We have explored whether there is a difference in the number of wide
binaries relative to close binaries as we move from the dense inner
cluster regions to the outer reaches of the cluster.  Such a
difference has been searched for but not found in earlier studies
(K\"ohler et al. 2006).  In view of the dramatic change in the
separation distribution function at 0$\farcs$5, we have chosen this
separation as a dividing point, such that we consider binaries between
0$\farcs$15 and 0$\farcs$5 as `close' and binaries between 0$\farcs$5
and 1$\farcs$5 as `wide'.  Figure~8 shows the cumulative distributions
of these close and wide binaries as function of distance from
$\theta^1$~Ori~C in the Trapezium. It is evident from the figure that
the two sets of binaries do not have the same radial distribution with
increasing distance from the cluster core. To explore this further, we
show in Figure~9 the ratio of wide to close binaries as a continuous
cumulative distribution, with the first point calculated 30$''$
outside the exclusion zone, that is, at a distance of 90$''$ from
$\theta^1$~Ori~C. For each distance, the curve gives the ratio for all
binaries from the edge of the exclusion zone to that particular
distance. In other words, as the curve moves away from the Trapezium
it accounts for more and more binaries, until the last points on the
curve represent the mean ratio of wide-to-close binaries for the
entire ONC. The dashed lines indicate the 1$\sigma$ errors on the
numbers. We have furthermore calculated the same ratio for the
Duquennoy \& Mayor (1991) binaries, and the two dotted lines indicate
the values calculated from their Gaussian fit (lower line) and their
actual data points (upper line). These lines thus represent the ratio
for field stars.

The curve does not take into account that 9 of the binaries are likely
to be line-of-sight associations. Given that the probability of being
a line-of-sight pair increases with separation and with proximity to
the center of the ONC, it follows that the curve would dip even
deeper down in the inner region if we could remove the 9 non-physical
pairs.

It is evident that there is a very pronounced and almost monotonic
change in the ratio of wide to close binaries as one moves away from
the core of the ONC until a distance of about 460$''$, at which
point the ratio becomes flat. It is also clear that the mean ratio of
wide to close binaries for the whole ONC is lower than the Duquennoy
\& Mayor values.  We discuss the implications of these findings in
Section~5.

\subsection{Higher-order Multiples} 

In addition to the 72 binaries in the separation range from 0$\farcs$15
to 1$\farcs$5, we have found 3 triple systems with separations in the
same range.  Tokovinin (2001) suggests that the ratio of triples to
binaries is 0.11$\pm$0.04 in general, whereas after subtracting the 9
line-of-sight pairs we find 0.048$\pm$0.01 in the ONC. Within the
considerable errors the numbers are close to being consistent, but it should
also be recalled that our survey is restricted to the limited range
between the 0$\farcs$15 completeness limit and the 1$\farcs$5 confusion
limit. It is entirely possible that some close pairs in triple systems
have already evolved to become spectroscopic binaries and have thus
become unobservable with our detection method. Nor can it be excluded
that some wide pairs exist in triple systems with separations larger
than 1$\farcs$5, although such wide systems cannot be common (Scally et
al. 1999).  Much better statistics is required to settle the question
whether the ONC might be deficient in triple systems.

\subsection{Flux Ratios and the Nature of Companions} 

We have determined the flux-ratios of those of our binaries that are
not saturated in our images. Figure~10 shows the resulting histogram
as a function of $\Delta$mag. As is well established from other binary
studies, the majority of binaries in the ONC also have unequal
components, although the bin with equal components ($\Delta$m $<$
0.5~mag.) is the most populated. Only very few of the companions have
a magnitude difference to their primary of more than 2.5 magnitudes.

In the absence of spectroscopic information, we have made a crude
attempt to investigate the nature of the companions based on their
measured flux ratios. In order to do that we have made some
assumptions. First, we assume that the observed flux ratios reflect
the photospheric fluxes, in other words that H$\alpha$ emission line
fluxes are not seriously affecting the ratios. This may not always be
a good assumption, since mid-to-late M dwarfs often are very active,
and their photospheric fluxes are so low that H$\alpha$ line emission
could be a significant contribution. If we mistake H$\alpha$ line
emission for photospheric flux from the primary, our estimate of a
spectral type for a companion will be earlier than it is in
reality. Second, most published photometry of late-type dwarfs is
broad-band, whereas we have observed in the narrow-band H$\alpha$
filter, so we assume that our observed magnitude differences can be
compared to R-band photometry. Since we are dealing with the
difference between two stars, this is probably not a bad assumption,
at least when the flux ratio is not large. Third, we assume that the
observed flux ratios are not affected by differences in extinction
between the components. Given the young age and occasional association
with molecular clouds, this may not in all cases hold true.

With these caveats spelled out, we have used the M$_I$ vs spectral
type relation and the R-I colors for M dwarfs (e.g., Dahn et al. 2002)
to derive the difference in R-band magnitudes as a function of
spectral type for M-dwarfs. The relation turns out to be essentially
linear, with a mean drop of 0.56 magnitudes per spectral subtype
throughout the M spectral range. We then used the spectral
classifications for the primaries provided by SIMBAD, of mixed
provenance, with our flux ratios to estimate a spectral type for the
secondaries. For a cluster with an age between 0.5 and 3 Myr, the
substellar limit is around spectral type M6.25 following the models of
Baraffe et al. (1998) and Chabrier et al. (2000) and using the
temperature scale of Luhman et al. (2003), see also Luhman et
al. (2006). To our surprise, quite a number of the secondaries appear
to be brown dwarfs, in at least one case forming a wide BD-BD
binary. We comment on selected cases in the Appendix. Given the
assumptions involved and the simplistic nature of these spectral type
estimates, it is obvious that spectroscopy is required to establish
the true nature of the secondaries.

\section{DISCUSSION}

\subsection{Structure and Evolution of the Separation Distribution Function}

The separation distribution function of field binaries has been
approximated by various functions. \"Opik (1924) suggested that
binaries with separations larger than $\sim$60~AU follow a {\em f(a)
$\sim$ 1/a} distribution, whereas Kuiper (1942) proposed that the
overall distribution could be represented by a Gaussian. The latter
distribution was adopted as a good fit to their data in the
influential study by Duquennoy \& Mayor (1991), although it is evident
from Figure~7 that in the specific separation range discussed in the
present paper, the Gaussian is a poor approximation to the actual
data. Integrating the \"Opik relation over logarithmic bins results in
a straight horizontal line to describe the logarithmic separation
distribution, and this is clearly a better fit to the two data points
of Duquennoy \& Mayor seen in Figure~7.  The \"Opik relation has also
been supported by numerous studies summarized by Poveda \& Allen
(2004).

For the separation distribution of ONC binaries (corrected for
line-of-sight associations) both a Gaussian and a horizontal line
provide acceptable fits to the data when considering the uncertainties
of the data points. It is thus not possible to classify the structure
of the ONC separation distribution function with the available data.

The separation distribution function in the ONC most probably has not
yet found its final shape. Two basic mechanisms operate that can
affect the orbit of a young binary: rapid dynamical decay in small-N
clusters (e.g., Sterzik \& Durisen 1998, Reipurth \& Clarke 2001, Bate
et al. 2002) and the passage of a binary through a dense cluster
(e.g., Kroupa 1995a,b, Kroupa et al. 1999). The former occurs
primarily during the Class~0 phase (Reipurth 2000), and is no longer
relevant for stars in the ONC. The latter, however, may be essential
for understanding the binary population of the ONC.  A wide binary
falling through the potential well of a cluster will gain kinetic
energy through encounters, and binaries with weak binding energies are
eventually disrupted (Heggie 1975). A prediction of this scenario is
that binaries at distances from the cluster center larger than the
corresponding crossing time should not be showing any dynamical
alterations due to encounters with other cluster members (Kroupa et
al. 1999, 2001).  The crossing time of a star through a cluster is
$t_{cross} = 2R/\sigma$, where $R$ is the cluster radius and $\sigma$
is the mean one-dimensional velocity dispersion in the
cluster. Assuming that this velocity dispersion in the ONC is of the
order of 2 km~s$^{-1}$ (e.g., Jones \& Walker 1988), we have in
Figure~9 indicated the crossing time for different distances to the
cluster center. The figure suggests that for distances larger than
roughly 460~arcsec there is no longer a measurable change in the ratio
of wide-to-close binaries.  Given that the wide-to-close binary ratio
changes by a factor of 4-5 from the inner to the outer regions, this
suggests that many, and perhaps most, of the wide binaries are
disrupted after only a few passages of the cluster center.  The
variation of the ratio of wide-to-close binaries from the inner to the
outer regions of the ONC, seen in Figure~9, offers the first
compelling observational evidence that dynamical interactions in the
dense central region of the ONC have taken place.

In principle, the diagram in Figure~9 allows us to determine the age
of the ONC.  An angular distance of $\sim$460~arcsec corresponds to a
crossing time of about 1 million years.  However, an age determined in
this manner is directly dependent on the velocity dispersion
assumed. Therefore, all we can say about the age of 1 million years we
estimate for the ONC is that it is consistent with other ONC age
estimates (e.g., Hillenbrand 1997).

The significantly lower number of binaries that we find in the ONC
compared to associations thus appears to be due, at least in part, 
dissolution of wide binaries. However, under certain circumstances an
encounter could lead to hardening of the binary, making it closer than
our resolution limit, so it is not lost to the overall binary budget.

Even in its outermost regions, the ONC shows a smaller ratio of
wide-to-close binaries than seen in the field by Duquennoy \& Mayor
(1991). Considering that the majority of field stars are likely to
have been formed in a cluster, it follows that many of the {\em wide}
binaries in the field must have formed in the gentler environment
of a loose T association. 

Durisen \& Sterzik (1994) found, on theoretical grounds, that binaries
are more likely to form in clouds with lower temperatures. Reipurth \&
Zinnecker (1993) found observational evidence that clouds with more
stars have relatively fewer binaries in the separation range under
study (mostly wide visual binaries). Both these results could indicate
that loose T associations produce or retain more wide binaries than do
clusters. However, Brandeker et al. (2006) noted that the young sparse $\eta$
Chamaeleontis cluster has a deficit of wide binaries. Unless this
small cluster is the remnant of a much denser cluster, then this
result would seem to be in contradiction to the notion that wide
binaries are preferentially formed/preserved in loose associations.

In any case, there is no question that the field binary population is
a mix of binaries formed in clusters and in associations.  We can
attempt to calculate the fraction of stars that originate in clusters
and in associations. In Sect. 4.2 we showed that, in the limited
separation range 67.5 - 675~AU, the binary fraction of the field stars
of Duquennoy \& Mayor (1991) is about a factor 1.5 higher than in the
ONC.  We also found that in the range 150 - 675~AU associations have a
factor 2.2 more binaries than the ONC. In other words, in these
limited ranges associations have about 1.5 times as many binaries as
the field population, and the ONC has only about 2/3 as many binaries
as the field population. It follows that the field star binary
population in principle can be produced if only 1/3 of binaries come from the
binary-rich associations and 2/3 originate in binary-poor ONC-type
clusters, in excellent agreement with the results of Patience et
al. (2002) and in approximate agreement with findings that 70-90\% of
all stars may be formed in clusters (e.g., Lada \& Lada 2003). Of
course, this assumes that the binary ratios between field stars,
associations, and the ONC derived here within narrow separation ranges
can be extrapolated to all separations, an assumption that may not
hold at all.

\subsection{Very Low Mass Stellar and Substellar Companions}

As noted in Sect.~4.5, spectroscopy is required to determine the true
nature of the companions we have found.  However, unless significant
extinction differences are common among our binaries, then the
difference in component brightness, combined with spectral type
estimates for the primaries when available, are indicative of the
spectral types of the secondaries. While this is highly uncertain in
individual cases, overall it is likely to be indicative of the
properties of the companion population. Table~1 lists the spectral
types for 46 primaries, and of these 38, or 83\%, are M-type stars.
We find that more than half (25) of these 46 binary stars have
secondaries with spectral types of M5 or later, and about one third
(17) could be substellar. These numbers are upper limits, since we
cannot correct for the 9 false binaries that we expect to be present
in our sample of 78 multiple systems, but they are probably more
likely to be found among pairs with faint, widely separated
companions. However, even if we subtracted all 9 objects from the 17
that could be substellar, this would still indicate that 8, or about
17\%, of the 46 binaries with spectral type estimates are likely to
have substellar companions. Selected cases are discussed in the
Appendix.

The closest binaries we were able to resolve with our imaging
technique have projected separations of 50~AU. All of the binaries
with likely substellar companions are thus quite wide, which is of
interest for formation theories (see, e.g., Lucas et al. 2005, and
reviews by Burgasser et al. 2007, Luhman et al. 2007, Whitworth et
al. 2007).

It is likely that there are several brown dwarf -- brown dwarf
binaries in our binary sample, but the only one that we are certain of
is COUP~1061. Infrared spectroscopy of this source indicates a
spectral type of M9-L0 (Meeus \& McCaughrean 2005). We have resolved
this object into two components with almost equal brightness and a
projected separation of 100~AU. Brown dwarf binaries with such large
separations are rare (e.g., Lucas et al. 2005, Allen et al. 2007),
although a few very wide pairs are known (e.g., Artigau et al. 2007,
Barrado y Navascu\'es et al. 2007).

\section{CONCLUSIONS}

We have analyzed a large set of H$\alpha$ images of the Orion Nebula
Cluster acquired with the Hubble Space Telescope and the Advanced
Camera for Surveys with the goal of detecting binaries among a sample
of 781 ONC members. The following results were obtained:

1. A total of 75 binaries and 3 triple systems were detected, of which
   55 are new discoveries. 

2. Within the limited angular range of 0$\farcs$15 to 1$\farcs$5,
   corresponding to projected separations of 67.5~AU to 675~AU, we
   have found a binary fraction of 8.8\%$\pm$1.1\% after correcting
   for a statistically determined contamination of 9 line-of-sight
   binaries.

3. The field binary fraction for solar type stars in the same
   separation range is 1.5 times larger, and for T Tauri associations
   it is 2.2 times larger than in the ONC, confirming earlier results
   that the ONC is deficient in binaries, now with statistically
   significant data.

4. The separation distribution function for young binaries in the ONC
   shows a dramatic decrease in binaries for angular separations
   larger than 0.5$''$, corresponding to projected separations of
   225~AU. 

5. The ratio of cumulative distributions of wide (0$\farcs$5 to 1$\farcs$5)
   to close (0$\farcs$15 to 0$\farcs$5) binaries show an increase out to a
   distance of about 460$''$ from the center of the ONC, after which
   it levels out. We interpret this as clear observational evidence
   for dynamical evolution of the binary population as a result of
   passages through the potential well of the ONC. These results are
   consistent with an age of the ONC of about 1 million yr.

6. It appears that much of the deficiency of binaries in the ONC
   compared to the field star population can be understood, at least
   in part, in terms of the destruction of wide binaries combined
   with a secondary effect from orbital evolution of binaries towards
   closer separations that are unobservable in direct imaging surveys.

7. Limited spectral information about the primary stars indicate that
   they are low mass T~Tauri stars (except for one Herbig Ae/Be
   star). Assuming that most binaries are not affected by differential
   extinction, we find that possibly as many as 50\% of the binaries
   have companions with spectral types later than M5, and from 1/6 to
   1/3 of the binaries may have substellar companions, all of which with
   separations of at least 50~AU. This large number of wide substellar
   companions is of interest for theories of brown dwarf formation.

\acknowledgments

We thank Rainer K\"ohler, Pavel Kroupa, Luiz Paulo Vaz, and Adam
Burgasser for valuable comments, and an anonymous referee for
a very helpful referee report.  We are grateful to Burton Jones for
providing us an electronic version of the Jones \& Walker (1988)
proper motion table. Marcelo Guimar\~aes acknowledges financial
support by CAPES, CNPq and FAPEMIG.  Based on observations taken under
program GO-9825 with the NASA/ESA Hubble Space Telescope obtained at
the Space Telescope Science Institute, which is operated by the
Association of Universities for Research in Astronomy, Inc., under
NASA contract NAS5-26555.  This material is based upon work supported
by the National Aeronautics and Space Administration through the NASA
Astrobiology Institute under Cooperative Agreement No. NNA04CC08A
issued through the Office of Space Science. This research has made use
of the SIMBAD database, operated at CDS, Strasbourg, France, and
NASA's Astrophysics Data System Bibliographic Services.





\appendix

\section{Appendix. Individual Binaries of Interest.}

In the following, comments are provided on binaries of particular
interest, especially regarding possible substellar companions.

{\bf V1438 Ori}~(JW 39). The spectral type of M3 is due to
Hillenbrand (1997). If the 2.6 mag brightness difference is not
affected by extinction differences, then the companion is an M8 brown
dwarf at a projected separation of 170~AU.  Stassun et al. (1999)
detected lithium in the primary.

{\bf JW 71}. The spectral type of M4 is due to Hillenbrand (1997). If
the 1.4 mag brightness difference is not affected by extinction
differences, then the companion is an M7 brown dwarf with a projected
separation of 90~AU.

{\bf V1118 Ori}~(JW 73). This is a member of the class of EXor's (Herbig
1989). The star, which is also known as Chanal's Object, has had 5
outbursts since its discovery in 1984, and typically varies within the range
14.5 $<$V$<$18, with a rise time of less than a year and a declining
phase about twice as long. Optical and infrared spectroscopy of
V1118~Ori is reported by Parsamian et al. (2002) and Lorenzetti et
al. (2006), and X-ray observations are analyzed by Audard et
al. (2005). The discovery that V1118~Ori has a companion only 0.4
magnitude fainter in the H$\alpha$ filter raises interesting questions
about which of the components that may drive the variability (Herbig
2007).

{\bf JW 121}.  The spectral type of M3 is due to Hillenbrand (1997). If
the 3.0 mag brightness difference is not affected by extinction
differences, then the companion is an M8 brown dwarf with a projected
separation of 153~AU.

{\bf JW 147}.  The spectral type of M5 is due to Hillenbrand (1997). If
the 1.4 mag brightness difference is not affected by extinction
differences, then the companion is an M8 brown dwarf with a projected
separation of 135~AU.

{\bf JW 152}.  The spectral type of M3 is due to Hillenbrand
(1997). If the 2.5 mag brightness difference is not affected by
extinction differences, then the companion is an M7 brown dwarf with a
projected separation of 81~AU.

{\bf JW 235}. JW 235 is associated with the HH 504 object, discovered
by Bally \& Reipurth (2001), see also O'Dell (2001).  The binarity of
the star was discovered by K\"ohler et al. (2006). Its proper motion in
the Jones \& Walker catalog implies that it has 0\% probability of
being a member of the ONC.  However, given the similar brightness of
the two components, it is likely that brightness variations of the
stars shifted the photocenter used for astrometry. Evidence for ONC
membership is strong and is based on the presence of H$\alpha$
emission, detection in X-rays, optical variability, and association
with an HH object.

{\bf V1274 Ori}~(JW 248). The spectral type of the primary is somewhat
in dispute. Hillenbrand (1997) suggests M0.5-M2 (with the Ca II lines
in emission), Edwards et al. (1993) suggest M3, and Meeus \&
McCaughrean (2005) suggest M0-M4. The binarity of the system was found
by the COUP survey.  Our primary is the secondary in the COUP catalog,
indicating that our secondary is highly X-ray active. Sicilia-Aguilar
et al. (2005) detected lithium in the (optical) primary, and found an
inverse P Cygni profile at H$\alpha$.  If there is not an extinction
difference between the primary and secondary, then the large flux
difference ($>$3.7 mag) suggests that the companion could be an L0
brown dwarf. However, the system is only 3 arcmin from
$\theta^1$~Ori~C, so there is a 1\% chance of a line-of-sight
association.

{\bf JW 296}. JW 296 is the primary of a hierarchical triple
system. The secondary and tertiary are very faint stars, surrounded by
a common proplyd-like envelope known as 066-652 (O'Dell \& Wong 1996). The
primary is of spectral type M4.5 according to Hillenbrand (1997). If
the brightness difference is taken at face value, the secondary and
tertiary should be L0 brown dwarfs, but given the obvious association
with nebulosity it is likely that their faintness is merely due to
extinction.

{\bf JW 355}.  The primary of this system is located at the edge of a
proplyd known as 109-246 (O'Dell \& Wong 1996). The proper motion of the star
suggests it is not an ONC member, but it is found to be a member by
the COUP project. Hillenbrand (1997) suggests a mid-K spectral
type. The system is only 90$''$ from $\theta^1$~Ori~C, and the star
density is so high that the possibility of a line-of-sight association
is several percent.

{\bf JW 370}. There is a silhouette disk close ($\sim$3$\farcs$4) to
this binary. Given the local density of stars around this system, the
probability that such a silhouette disk is just a chance alignment is
$\sim$7\%.  The chance of a line-of-sight alignment for the binary
itself is 1.5\%. Hillenbrand (1997) suggests a spectral type of K0-K3.

{\bf V1492 Ori}~(JW 383). The primary is of spectral type M3 according
to Hillenbrand (1997).  If the 2.1 mag brightness difference is not
affected by extinction differences, then the companion is an M7 brown
dwarf with a projected separation of 86~AU. Stassun et al. (1999)
derive a rotation period of 7.11 days for the primary, and notes the
presence of lithium in the spectrum.

{\bf Parenago 1806}~(JW 391). The primary is of spectral type M1 according to
Edwards et al. (1993) and Hillenbrand (1997), who also notes the
presence of Ca~II emission. The primary is saturated in our images,
but with a magnitude difference of at least 3.3 mag, the companion
would be an M7 brown dwarf if there is no difference in
extinction. However, the pair is located in a region of high stellar
density, so the propability of a chance alignment is more than 5\%.

{\bf JW 509}. This binary was first detected by Prosser et al. (1994)
and subsequently by Padgett et al. (1997) and Lucas et
al. (2005). Although it is a well known binary it does not have a
spectral type in the literature. Since this system is located in a
crowded region, the probability of a chance alignment is $\sim$ 2.7\%,
but the reality of the binary is supported by an interaction zone
between the stars, visible in our image (Fig. 3b).

{\bf Parenago 1914}~(JW 551). The primary of this hierarchical triple system
has a spectral type M1 according to Hillenbrand (1997). The tertiary
is not detectable either in the optical or in X-rays, and since its
separation from the primary is as large as 1$\farcs$27 it could be a
background object. The system seems to be part of a small subcluster,
which gives it a high probability of a chance alignment ($\sim$
2.2\%).  Given the primary spectral type, the difference in magnitudes
to the tertiary (3.6) and assuming that the tertiary is not a
background object, the tertiary could have a spectral type M7 or M8
and thus would be a brown dwarf candidate.

{\bf COUP 967}. This star was catalogued as a binary by Prosser et
al. (1994), Padgett et al. (1997) and Lucas et al. (2005).  Our image
shows that the primary is associated with a proplyd (184-427 in O'Dell
\& Wong 1996) and the secondary is a faint object ($\Delta$m$_{H\alpha}$
= 2.4). The spectral type for the primary is M2.5 according to
Hillenbrand (1997).  The system is very close to $\theta^1$~Ori~C
($\sim$ 70$''$) and the probability of a chance alignment is 3.5\%. If
the components have the same extinction, then the secondary would be
of spectral type M6.5, and thus at the hydrogen burning limit.

{\bf JW 592}. The spectral type M2.5 is due to Hillenbrand (1997) and
M4 according to Edwards et al. (1993).  The brightness difference (4
magnitudes) would imply an M9.5 spectral type or later for the
secondary, making it a brown dwarf with a projected separation of
$\sim$ 275 AU.

{\bf COUP 1061}. This very close system ($\sim$ 100 AU) has a combined
spectral type M9-L0 according to Meeus \& McCaughrean (2005), who did
not resolve it, and since the components have virtually the same
brightness, this is therefore a {\em bona fide} brown dwarf - brown
dwarf binary with a projected separation of 100~AU.

{\bf V1524 Ori}~(JW 681). This binary was first catalogued by Prosser
et al. (1994) and has a spectral type K7 according to Prosser \&
Stauffer (unpublished).  The secondary is associated with the proplyd
213-346 (O'Dell \& Wong 1996). In the Hillenbrand (1997) catalog this
system has two entries, both with number 681 but with different
coordinates.  SIMBAD identifies three different objects: H97b-681,
H97b-681a, and H97b-681b. The first is JW~681, also known as
V1524~Ori, and is the primary in this double. The second is the
unrelated object MLLA~312, and the third is the X-ray source
COUP~1149, which forms the secondary component, associated with the
proplyd.  Because this system is located near to $\theta^1$ Ori C
($\sim 77''$) the probability of a chance alignment is 3.5\%.

{\bf V1528 Ori}~(JW 727). The spectral type M2 is due to Hillenbrand
(1997) and the difference in brightness suggests a spectral type of L0
for the companion if the components have the same extinction. If so,
the secondary is a brown dwarf with projected separation of 329
AU. The primary shows H$\alpha$ emission according to our unpublished
survey.

{\bf JW 748}. This binary was discovered by Prosser et al. (1994), and
the spectral type of the primary is M3 according to Hillenbrand
(1997). Given the brightness diffference and if the components have
the same extinction, the secondary could be a brown dwarf with
spectral type M7 and a projected separation of 158~AU.

{\bf JW 767}. First discovered by K\"ohler et al. (2006) this system
has a spectral type M2.5 due to Hillenbrand (1997). The secondary has
a silhoutte disk and is associated with the HH~668 object (Bally et
al. 2006). The extinction caused by the silhoutte disk probably is
responsible for the large difference in brightness by more than 4
magnitudes between the primary and secondary.

{\bf Parenago 2075}~(JW 867). This object was observed by K\"ohler et
al. (2006) but despite its brightness and separation (0$\farcs$29)
they did not resolve it, suggesting possible major variability of the
secondary. The system presents evidence of H$\alpha$ emission, X-rays
and optical variability.  A spectral type of M1 was suggested by
Blanco (1963), and more recently Duncan (1993) assigned a spectral
type of K8Ve. Sicilia-Aguilar et al. (2005) reported lithium in the
spectrum of the primary.

{\bf JW 906}. The primary is of spectral type M3 according to
Hillenbrand (1997), and if the difference in brightness (2.7 mag) is
not affected by extinction then the secondary is a brown dwarf with
spectral type M8 and a large projected separation of 625
AU. Sicilia-Aguilar et al. (2005) reported lithium in the spectrum of
the primary.

\clearpage

\begin{deluxetable}{ccccccccccccccclc}
\tabletypesize{\scriptsize}
\tablewidth{500pt}
\tablecaption{Pre-main sequence binaries with angular separation between 0$\farcs$15 and 1$\farcs$5, 
outside the $60^{\prime\prime}$ exclusion zone around $\theta^1~$Ori C.}
\tablehead{
\colhead{JW} & \colhead{\%} & \colhead{Par} & \colhead{GCVS} & \colhead{COUP} & \colhead{H97} & \colhead{RA$_{2000}$} & \colhead{DEC$_{2000}$} & \colhead{PA} & \colhead{SEP [$''$]} & 
\colhead {m$_I$} & \colhead{$\Delta$m$_{H\alpha}$\tablenotemark{1}} & \colhead{PA$_{OriC}$} & \colhead{OriC} & \colhead{SpT\tablenotemark{2}} & \colhead{Memb.\tablenotemark{3}} & \colhead{Bin} }
\startdata
 39 & 99 & \nodata   & V1438   & \nodata &  39 & 5:34:38.1 & -5:27:41 &  21 & 0.38 & 14.29 & 2.6 & 246 & 628 & M3 & PV &   \\
 52 & 98 & \nodata   & \nodata & \nodata &  52 & 5:34:40.8 & -5:28:09 & 231 & 0.20 & 14.58 & 0.1 & 242 & 604 & M5 & P &   \\
 63 & 99 &     1569  & V1441   &      17 &  63 & 5:34:43.0 & -5:20:07 &   0 & 0.27 & 12.75 & 1.5 & 291 & 537 & K6 & PXV &   \\
 71 & 99 & \nodata   & \nodata &      21 &  71 & 5:34:44.5 & -5:24:38 & 105 & 0.20 & 15.61 & 1.4 & 261 & 483 & M4 & PX &   \\
 73 & 99 & \nodata   & V1118   & \nodata &  73 & 5:34:44.7 & -5:33:42 & 329 & 0.18 & 14.04 & 0.4 & 217 & 779 & M1 & PHV &   \\
 81 & 97 &     1600  & \nodata &      28 &  81 & 5:34:46.4 & -5:24:32 & 272 & 1.34 & 12.91 & 1.9 & 261 & 454 & M0 & PXV &   \\
121 & 99 &   \nodata & \nodata & \nodata & 121 & 5:34:51.2 & -5:16:55 & 200 & 0.34 & 14.90 & 3.0 & 316 & 541 & M3 & PV &   \\
124 & 99 &   \nodata & \nodata &      64 & 124 & 5:34:51.8 & -5:21:39 & 200 & 0.48 & 13.94 & 0.0 & 286 & 382 & M3.5 & PXV &   \\
127 & 99 &   \nodata & \nodata &      66 & 127 & 5:34:52.1 & -5:24:43 & 254 & 0.48 & 14.28 & 0.6 & 258 & 373 & M3.5 & PXHV &   \\
128 & 97 &   \nodata & V1458   &      67 & 128 & 5:34:52.2 & -5:22:32 & 215 & 0.39 & 12.64 & $>$0.6 & 278 & 366 & M2.5 & PXHV &   \\
135 & 99 &   \nodata & V1460   &      72 & 135 & 5:34:52.7 & -5:29:46 &  11 & 0.40 & 14.45 & 0.1 & 223 & 522 & M3 & PXV &   \\
147 & 99 &   \nodata & \nodata & \nodata & 147 & 5:34:54.4 & -5:17:21 &  26 & 0.30 & 14.80 & 1.4 & 318 & 489 & M5 & P &   \\
151 & 99 &   \nodata & \nodata &      95 & 151 & 5:34:54.8 & -5:25:13 & 332 & 0.15 & 15.44 & 1.6 & 251 & 341 & M3 & PX &   \\
152 & 99 &   \nodata & \nodata &      96 & 152 & 5:34:55.1 & -5:25:30 &  51 & 0.18 & 14.12 & 2.5 & 248 & 343 & M3 & PX &   \\
176 & 99 &     1673  & KQ      &     123 & 176 & 5:34:57.8 & -5:23:53 & 336 & 1.28 & 11.58 & $>$1.1 & 264 & 280 & K8 & PXV & e \\
190 & 99 &   \nodata & \nodata &     134 & 190 & 5:34:59.3 & -5:23:33 & 355 & 0.56 & 15.33 & 0.3 & 268 & 256 & M6 & PX & f \\
201 & 99 &   \nodata & \nodata &     150 & 201 & 5:35:01.0 & -5:24:10 & 246 & 1.09 & 13.41 & 0.8 & 258 & 235 & M2.5 & PXH &   \\
222 & 98 &   \nodata & V1320   &     174 & 222 & 5:35:02.2 & -5:29:10 &  89 & 1.07 & 14.14 & 1.4 & 212 & 407 & M2 & PXV &   \\
223 & 99 &   \nodata & \nodata &     177 &223a & 5:35:02.4 & -5:20:47 & 130 & 0.34 & 13.64 & 1.3 & 307 & 261 & \nodata & PX &   \\
224 & 99 &   \nodata & \nodata &     180 & 224 & 5:35:02.7 & -5:19:45 & 164 & 0.27 & 15.38 & 2.4 & 317 & 299 & M1 & PXV &   \\
235 &  0 &   \nodata & \nodata &     197 & 235 & 5:35:03.6 & -5:29:27 & 346 & 0.54 & 13.69 & 0.4 & 208 & 411 & \nodata & PXHV & d \\
248 & 99 &   \nodata & V1274   &     214 & 248 & 5:35:04.4 & -5:23:14 & 320 & 0.90 & 12.73 & $>$3.7 & 273 & 180 & M3 & PXV & e \\
  - &  - &   \nodata & \nodata &     260 &3064 & 5:35:06.2 & -5:22:13 & 284 & 0.40 & 16.20 & 0.3 & 295 & 169 & M4 & X & f \\
296 & 99 &   \nodata & \nodata &     275 & 296 & 5:35:06.6 & -5:26:51 & 196 & 0.89 & 14.30 & 3.2 & 215 & 255 & M4.5 & PX &   \\
296 & 99 &   \nodata & \nodata &     275 & 296 & 5:35:06.6 & -5:26:52 & 158 & 0.27 & 14.30 & 0.4 & 215 & 256 & M4.5 & PX &   \\
305 & 99 &   \nodata & \nodata & \nodata & 305 & 5:35:07.6 & -5:24:01 & 302 & 0.45 & 14.58 & 0.3 & 254 & 137 & M3 & PV & f \\
355 &  0 &   \nodata & \nodata &     403 & 355 & 5:35:10.9 & -5:22:46 & 197 & 0.42 & 15.03 & 3.5 & 294 &  90 & K: & X &   \\
370 & 99 &   \nodata & \nodata &     452 & 370 & 5:35:11.9 & -5:19:26 & 122 & 0.73 & 13.86 & 3.5 & 344 & 246 & K1.5 & PX &   \\
383 & 99 &   \nodata & V1492   &     489 & 383 & 5:35:12.7 & -5:16:14 & 280 & 0.19 & 14.57 & 2.1 & 353 & 433 & M3 & PXV &   \\
392 & 99 &   \nodata & \nodata &     498 & 392 & 5:35:12.7 & -5:27:11 & 185 & 0.23 & 15.21 & 0.2 & 194 & 234 & M6 & PX &   \\
391 & 99 &     1806  & \nodata &     501 & 391 & 5:35:12.8 & -5:20:44 &  94 & 0.29 & 12.97 & $>$3.3 & 341 & 168 & M1 & PXV &   \\
399 & 99 &   \nodata & \nodata &     523 &399b & 5:35:13.2 & -5:22:21 & 345 & 0.22 & 16.77 & 0.1 & 322 &  78 & \nodata & PX & a,c \\
410 & 99 &   \nodata & \nodata & \nodata & 410 & 5:35:13.2 & -5:36:18 & 167 & 1.19 & 16.00 & 1.3 & 184 & 777 & \nodata & PV &   \\
406 & 99 &   \nodata & V1327   &     543 & 406 & 5:35:13.5 & -5:17:10 & 271 & 0.95 & 13.88 & 2.6 & 353 & 375 & M1 & PXV & d \\
  - &  - &   \nodata & \nodata &     562 &9048 & 5:35:13.6 & -5:21:21 &  62 & 0.24 & 16.27 & 1.5 & 341 & 129 & M3 & X & a \\
422 & 99 &   \nodata & V1495   &     566 & 422 & 5:35:13.7 & -5:28:46 &  74 & 0.31 & 14.43 & 0.0 & 187 & 326 & \nodata & PXV &   \\
436 &  0 &   \nodata & \nodata &     620 &436a & 5:35:14.3 & -5:22:04 & 238 & 0.32 & 16.97 & 1.0 & 338 &  85 & \nodata & X & a,c\\
439 & 99 &   \nodata & V1329   &     626 & 439 & 5:35:14.5 & -5:17:25 & 150 & 0.30 & 14.97 & 0.1 & 355 & 359 & M1 & PXV &   \\
439 & 99 &   \nodata & V1329   &     626 & 439 & 5:35:14.5 & -5:17:25 &  78 & 1.21 & 14.97 & 4.1 & 355 & 359 & M1 & PXV &   \\
444 & 99 &   \nodata & \nodata &     651 & 444 & 5:35:14.6 & -5:16:46 & 190 & 0.18 & 15.53 & 1.0 & 356 & 398 & \nodata & PX &   \\
445 & 26 &   \nodata & \nodata &     645 & 445a& 5:35:14.7 & -5:20:42 & 191 & 0.44 & 14.67 & 1.8 & 351 & 163 & \nodata & X & a \\
  - &  - &   \nodata & V1500   & \nodata & 466 & 5:35:14.9 & -5:36:39 &  82 & 0.38 & 14.45 & 2.2 & 182 & 797 & \nodata & V &   \\
465 & 97 &     1875  & V409    & \nodata & 465 & 5:35:14.9 & -5:38:06 & 271 & 0.43 & 14.71 & 4.1 & 182 & 883 & \nodata & PV &   \\
498 & 99 &     1873  & V1504   & \nodata & 498 & 5:35:15.8 & -5:32:59 & 122 & 0.70 & 13.81 & 0.2 & 181 & 576 & \nodata & PHV &   \\
509 & 99 &   \nodata & \nodata &     789 &509a & 5:35:16.2 & -5:24:56 &  46 & 0.49 & 15.06 & 0.1 & 182 &  93 & \nodata & PXV & a,b,f\\
511 & 99 &   \nodata & \nodata & \nodata &511a & 5:35:16.3 & -5:22:10 & 241 & 0.41 & 15.67 & 2.4 & 358 &  73 & M1 & PX & a,c,e,f \\
  - &  - &   \nodata & \nodata &     822 &3031 & 5:35:16.8 & -5:17:17 &  84 & 0.26 & 16.37 & 0.5 &   1 & 366 & \nodata & X &   \\
551 & 99 &     1914  & \nodata &     881 &551a & 5:35:17.4 & -5:25:45 & 265 & 0.14 & 13.87 & 0.0 & 174 & 142 & M1 & PX & a \\
551 & 99 &     1914  & \nodata &     881 &551a & 5:35:17.4 & -5:25:45 & 262 & 1.27 & 13.87 & 3.6 & 174 & 143 & M1 & PX &   \\
552 & 99 &     1908  & V410    &     897 &552a & 5:35:17.5 & -5:21:46 & 156 & 0.46 & 14.50 & $>$1.8 &   9 &  99 & \nodata & PXV & a,c \\
560 & 98 &   \nodata & V1334   &     927 & 560 & 5:35:17.9 & -5:15:33 & 235 & 0.60 & 14.17 & 3.2 &   3 & 471 & \nodata & PXV &   \\
570 & 99 &   \nodata & \nodata &     937 &570a & 5:35:17.9 & -5:25:34 &   0 & 0.15 & 14.73 & 0.8 & 170 & 133 & \nodata & PX & a,b \\
566 &  0 &   \nodata & \nodata &     939 & 566 & 5:35:18.0 & -5:16:13 & 214 & 0.86 & 14.93 & 0.0 &   3 & 430 & \nodata & X & d \\
  - &  - &   \nodata & \nodata &     967 &9224 & 5:35:18.4 & -5:24:27 &  78 & 0.42 & 15.16 & 2.4 & 155 &  70 & M2.5 & X & a,b,f \\
592 & 99 &   \nodata & \nodata &     974 & 592 & 5:35:18.5 & -5:18:21 & 285 & 0.61 & 14.37 & 4.0 &   6 & 304 & M2.5 & PX &   \\
597 &  0 &   \nodata & \nodata &     994 & 597 & 5:35:18.8 & -5:14:46 & 307 & 0.17 & 14.02 & 0.1 &   4 & 519 & \nodata & XHV &   \\
  - &  - &   \nodata & \nodata &     998 &9239 & 5:35:18.8 & -5:22:23 & 314 & 0.20 & 17.54 & 0.3 &  31 &  70 & \nodata & X & a \\
  - &  - &   \nodata & \nodata &    1061 &5066 & 5:35:20.0 & -5:18:47 &  71 & 0.22 & 17.31 & 0.1 &  11 & 281 & M9 & 0X00 &   \\
638 & 99 &   \nodata & \nodata &    1077 & 638 & 5:35:20.0 & -5:29:12 & 354 & 1.01 & 17.03 & 3.2 & 171 & 353 & \nodata & PXV &   \\
681 & 99 &   \nodata & V1524   & \nodata &681a & 5:35:21.4 & -5:23:45 & 232 & 1.09 & 16.39 & 0.9 & 107 &  77 & K7 & PX0V & a \\
687 & 54 &   \nodata & \nodata &    1158 &687a & 5:35:21.7 & -5:21:47 & 231 & 0.49 & 14.66 & 2.1 &  39 & 124 & \nodata & XV & a,c \\
709 & 99 &     1994  & \nodata &    1202 & 709 & 5:35:22.2 & -5:26:37 & 309 & 0.22 & 12.92 & 0.5 & 156 & 213 & M0.5 & PX &   \\
722 & 98 &   \nodata & \nodata &    1208 & 722 & 5:35:22.3 & -5:33:56 & 216 & 0.36 & 14.51 & 0.2 & 172 & 639 & M4.5 & PX &   \\
727 & 99 &   \nodata & V1528   &    1233 & 727 & 5:35:22.8 & -5:31:37 & 285 & 0.73 & 13.87 & 4.7 & 169 & 503 & M2 & PXHV &   \\
748 & 99 &   \nodata & \nodata &    1279 &748a & 5:35:24.1 & -5:21:33 & 277 & 0.35 & 14.77 & 2.0 &  46 & 159 & M3 & PXV & a \\
767 & 99 &   \nodata & \nodata &    1316 & 767 & 5:35:25.2 & -5:15:36 &  75 & 1.13 & 13.89 & $>$4.1 &  16 & 485 & M2.5 & PXV & d \\
776 & 99 &   \nodata & V496    &    1328 &776a & 5:35:25.4 & -5:21:52 &  73 & 0.50 & 14.20 & 1.8 &  56 & 162 & \nodata & PXV & a \\
777 & 80 &   \nodata & \nodata &    1327 & 777 & 5:35:25.5 & -5:21:36 & 347 & 1.19 & 15.15 & 1.2 &  52 & 173 & K6 & X &   \\
783 & 95 &   \nodata & \nodata & \nodata & 783 & 5:35:25.5 & -5:34:03 & 283 & 0.40 & 14.17 & 0.8 & 168 & 656 & \nodata & PX &   \\
797 & 99 &   \nodata & \nodata &    1363 & 797 & 5:35:26.6 & -5:17:53 & 323 & 1.42 & 16.30 & 2.8 &  25 & 363 & \nodata & PX &   \\
  - &  - &   \nodata & \nodata &    1425 &3042 & 5:35:29.5 & -5:18:46 & 329 & 0.23 & 16.06 & 2.6 &  35 & 338 & \nodata & XV &   \\
841 & 99 &   \nodata & \nodata & \nodata & 841 & 5:35:30.0 & -5:12:28 &  16 & 0.41 & 14.41 & 0.1 &  17 & 686 & M4 & PHV &   \\
867 & 99 &     2075  & \nodata &    1463 & 867 & 5:35:31.3 & -5:18:56 & 212 & 0.29 & 12.19 & $>$0.3 &  40 & 347 & K8 & PXHV &   \\
884 & 99 &   \nodata & \nodata & \nodata & 884 & 5:35:32.4 & -5:14:25 & 287 & 1.32 & 14.91 & 2.8 &  24 & 589 & \nodata & P &   \\
893 & 99 &   \nodata & \nodata & \nodata & 893 & 5:35:33.2 & -5:14:11 & 208 & 0.15 & 14.07 & 0.0 &  24 & 606 & \nodata & PV &   \\
906 & 99 &   \nodata & \nodata & \nodata & 906 & 5:35:34.7 & -5:34:38 & 293 & 1.39 & 14.17 & 2.7 & 158 & 728 & M3 & PV &   \\
924 & 99 &   \nodata & \nodata & \nodata & 924 & 5:35:36.5 & -5:34:19 &  73 & 1.04 & 16.63 & 1.3 & 156 & 722 & \nodata & P &   \\
945 & 99 &     2149  & \nodata & \nodata & 945 & 5:35:40.2 & -5:17:29 &  46 & 1.41 & 12.51 & 3.6 &  45 & 501 & B6 & P &   \\
\enddata
\tablenotetext{1}{The symbol $>$ indicates that the primary is saturated (sometimes both stars), thus $\Delta$m is only an approximation}
\tablenotetext{2}{Spectral types obtained at SIMBAD}
\tablenotetext{3}{Membership criteria used: P - proper motion catalog by Jones \& Walker (1988) with probability $\ge$ 93\%, X - COUP source by Getman et al. (2005), 
H - H$\alpha$ emission (Pettersson et al., in prep.), V - variability noted by the General Catalog of Variable Stars}
\begin{verse}
a - Prosser et al. (1994) \\
b - Padgett et al. (1997) \\
c - Simon et al. (1999) \\
d - Koehler et al. (2006) \\
e - Getman et al. (2005) \\
f - Lucas  et al. (2005) \\
\end{verse}
\end{deluxetable}

\begin{deluxetable}{ccccccccccccccclcc}
\tabletypesize{\scriptsize}
\tablewidth{525pt}
\tablecaption{Other binaries with angular separations $<$1$\farcs$5 toward the ONC.}
\tablehead{
\colhead{JW} & \colhead{\%} & \colhead{Par} & \colhead{GCVS} & \colhead{COUP} & \colhead{H97} & \colhead{RA$_{2000}$} & \colhead{DEC$_{2000}$} & \colhead{PA} & \colhead{SEP [$''$]} & 
\colhead{m$_I$} & \colhead{$\Delta$m$_{H\alpha}$} & \colhead{PA$_{OriC}$} & \colhead{OriC} & \colhead{SpT\tablenotemark{*}} & \colhead{Memb.} & \colhead{Bin} & \colhead{C}}
\startdata
553 & 99 &      1911 & V1510   &     899 &553a & 5:35:17.6 & -5:22:57 & 113 & 0.36 & 12.41 & 1.4 &  34 &  31 & K3.5 & PXV & a &1 \\
596 & 99 &      1927 & AF      &     986 & 596 & 5:35:18.7 & -5:23:14 &  78 & 0.30 & 13.06 & 2.2 &  75 &  34 & K3.5 & PXV &   &1 \\
\nodata &  \nodata &\nodata  & \nodata &1085 & 3075 & 5:35:20.2 & -5:23:09 & 104 & 0.21 & \nodata & 0.4 &  76 &  57 & \nodata & X & &1 \\
182 & 48 &   \nodata & \nodata &     127 & 182 & 5:34:58.0 & -5:29:41 &  77 & 0.11 & 16.56 & 0.2 & 216 & 467 & \nodata & X &   &2 \\
290 & 99 &   \nodata & \nodata &     266 & 290 & 5:35:06.4 & -5:27:05 &  84 & 0.11 & 15.48 & 1.3 & 214 & 268 & \nodata & PX &   &2 \\
  - &  - &   \nodata & \nodata &    1195 &3034 & 5:35:22.3 & -5:18:09 & 227 & 0.11 & 15.67 & 0.8 &  15 & 326 & \nodata & X &   &2 \\
\nodata &  \nodata &    1423 & \nodata &   \nodata &       3114 & 5:34:19.5 & -5:27:12 & 168 & 0.56 & 10.52 & 0.1 & 255 & 881 & G5 & & &3\\
     58 &        0 &\nodata  & \nodata &   \nodata &         58 & 5:34:41.8 & -5:34:30 & 282 & 1.43 & 16.13 & 2.2 & 218 & 844 & \nodata & & &3\\
     61 &       46 &\nodata  & \nodata &   \nodata &         61 & 5:34:42.7 & -5:28:37 & 153 & 0.45 & 13.52 & 0.9 & 238 & 594 & \nodata & & &3\\
\nodata &  \nodata &\nodata  & \nodata &   \nodata &    \nodata & 5:34:45.8 & -5:30:58 & 324 & 0.46 & \nodata & 0.1 & 225 & 645 & \nodata & & &3\\
\nodata &  \nodata &\nodata  & \nodata &   \nodata &      10343 & 5:35:01.4 & -5:24:13 & 292 & 0.28 & \nodata & 0.6 & 257 & 231 & \nodata & & &3\\
\nodata &  \nodata &\nodata  & \nodata &   \nodata &       9221 & 5:35:18.3 & -5:24:39 &  96 & 0.67 & 16.72 & 2.2 & 160 &  81 & \nodata & & &3\\
\nodata &  \nodata &\nodata  & \nodata &   \nodata &    \nodata & 5:35:30.0 & -5:34:31 & 237 & 1.29 & \nodata & 0.3 & 163 & 699 & \nodata & & &3\\
\enddata
\tablenotetext{*}{Spectral types obtained at SIMBAD}
\begin{verse}
a - Prosser et al. (1994) \\
1 - Binaries with angular separation $<$0$\farcs$4 inside the exclusion zone\\
2 - Binaries with angular separation $<$0$\farcs$15 outside the exclusion zone\\
3 - Binaries with angular separation $<$1$\farcs$5 outside the exclusion zone but with no evidence of ONC membership
\end{verse}
\end{deluxetable}

\clearpage

\begin{figure}
\epsscale{2.0}
\plotone{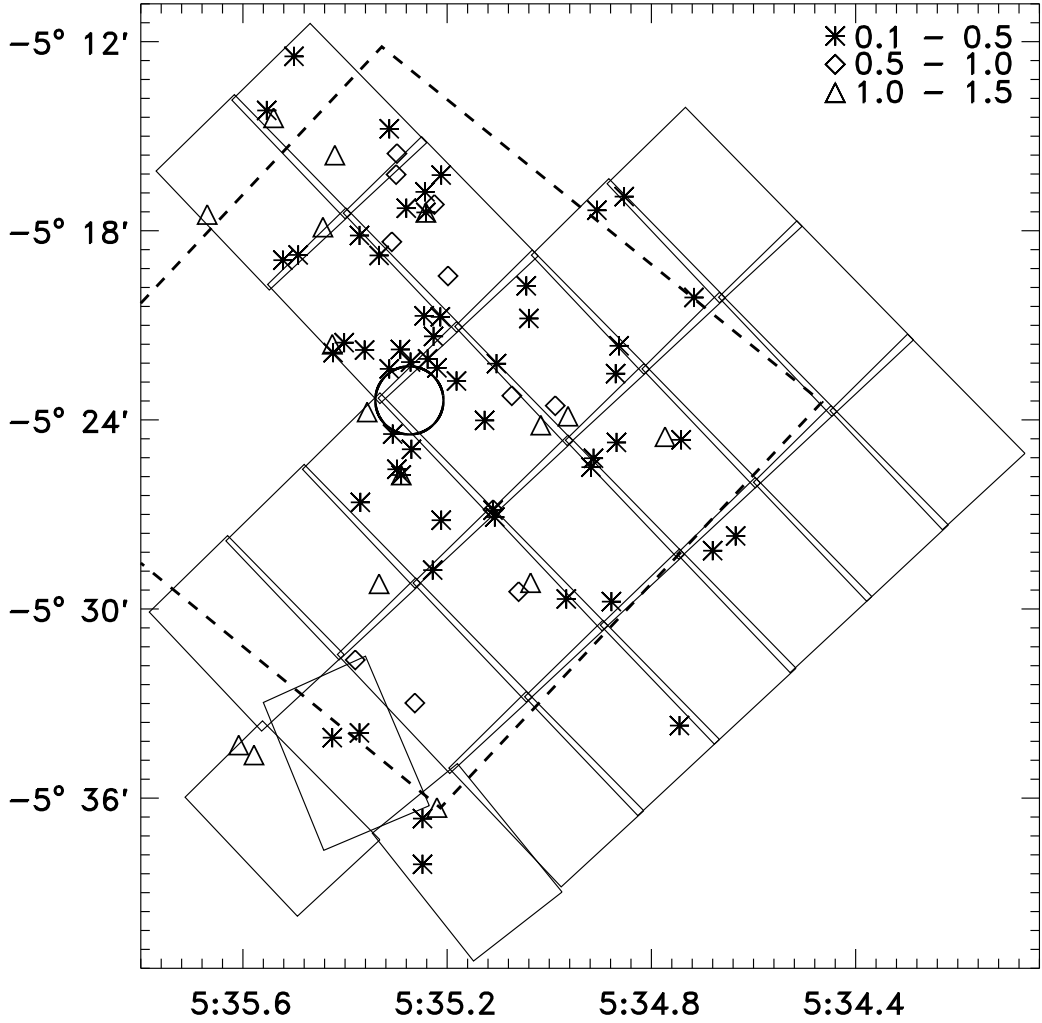}
\caption{Distribution of HST images across the ONC. The 60$''$
  exclusion zone centered on $\theta^1$~Ori~C is marked by a circle,
  and the boundaries of the COUP survey is indicated with dashed
  lines. Binaries in three different separation ranges are marked by
  different symbols (asterisks: 0$\farcs$1 - 0$\farcs$5, diamonds:
  0$\farcs$5 - 1$\farcs$0, triangles: 1$\farcs$0 - 1$\farcs$5).
  Coordinates are equinox 2000.
\label{fig1}}
\end{figure}

\clearpage

\begin{figure}[htb]
\epsscale{2.4}
\plotone{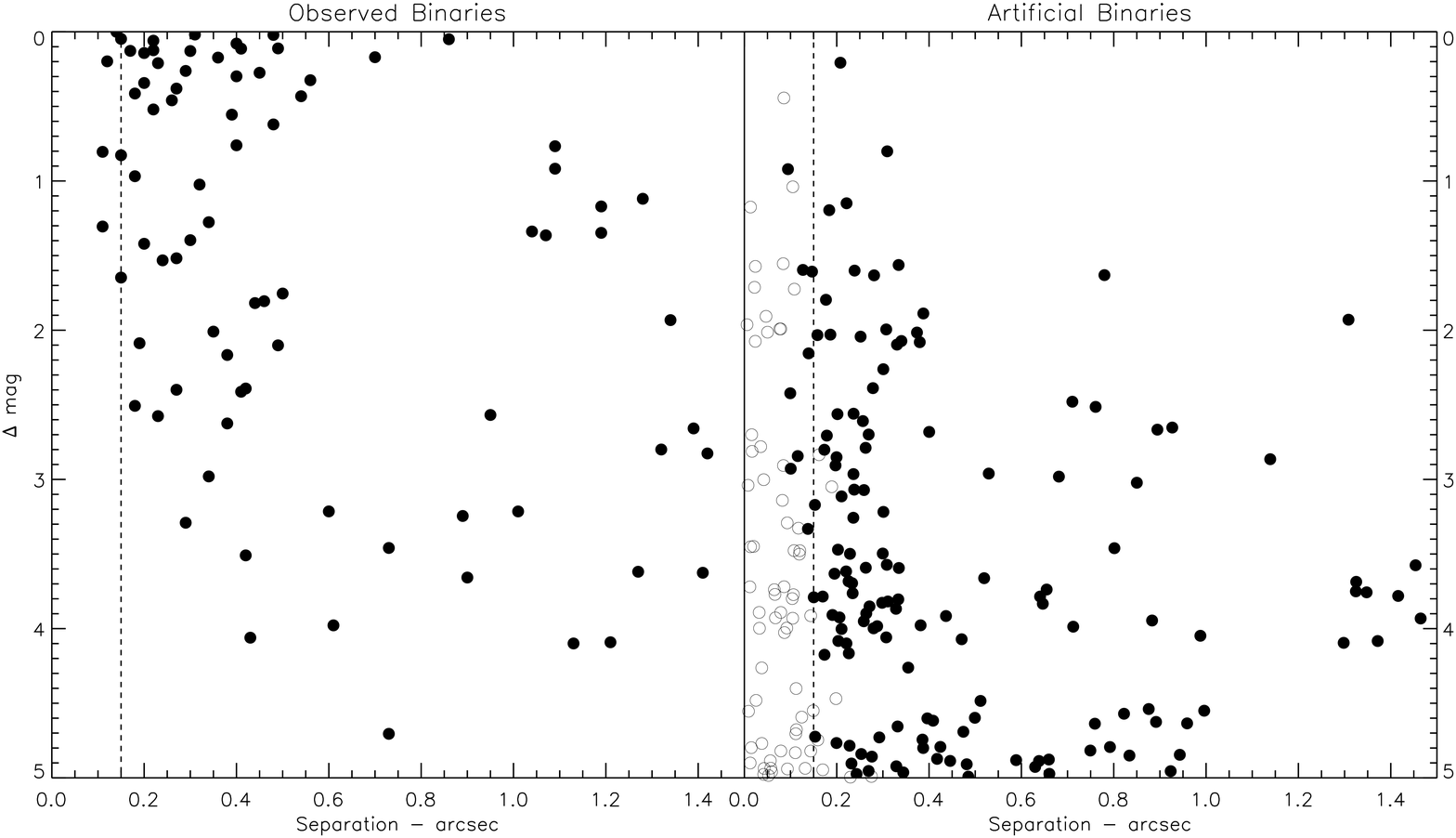}
\caption{Two plots showing $\Delta$m vs. binary separation in
  arcseconds. The left panel shows the actual binaries observed, while
  the right panel shows ``observations'' of actual stellar images to
  which artifical companions were added with random separations,
  position angles, and $\Delta$m. Filled circles indicate those
  companions we could detect, and empty circles those we failed to
  see. The vertical dashed line indicates the 0$\farcs$15 limit we
  adopted as our completeness limit.
\label{fig2}}
\end{figure}

\clearpage

\begin{figure}[htb]
\epsscale{2.0}
\plotone{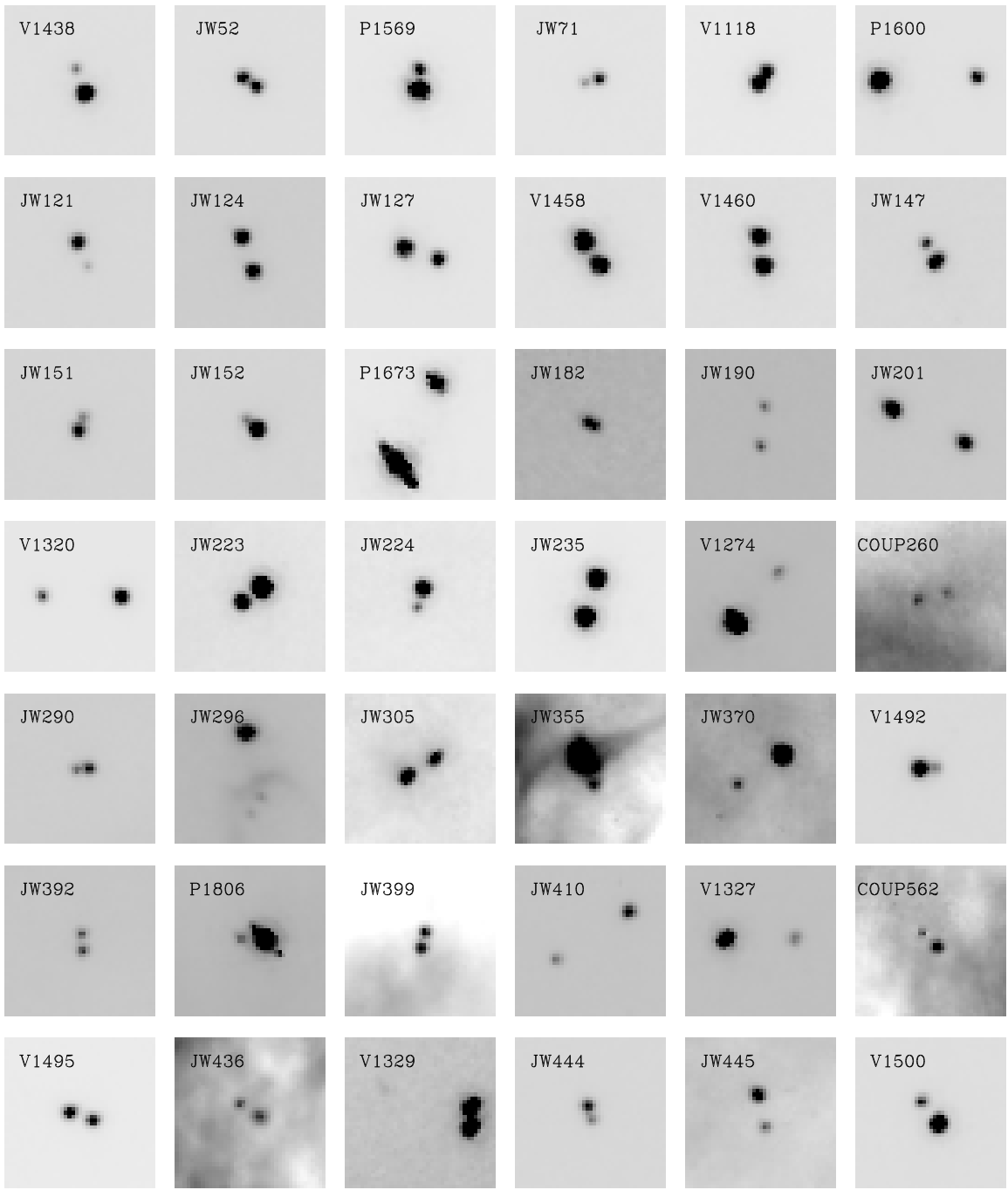}
\caption{All binaries identified among the ONC members. Each stamp is
2$''$ wide, and in each panel, north is up and east is to the left.
\label{fig3a}}
\end{figure}

\clearpage

\setcounter{figure}{2}

\begin{figure}[htb]
\epsscale{2.0}
\plotone{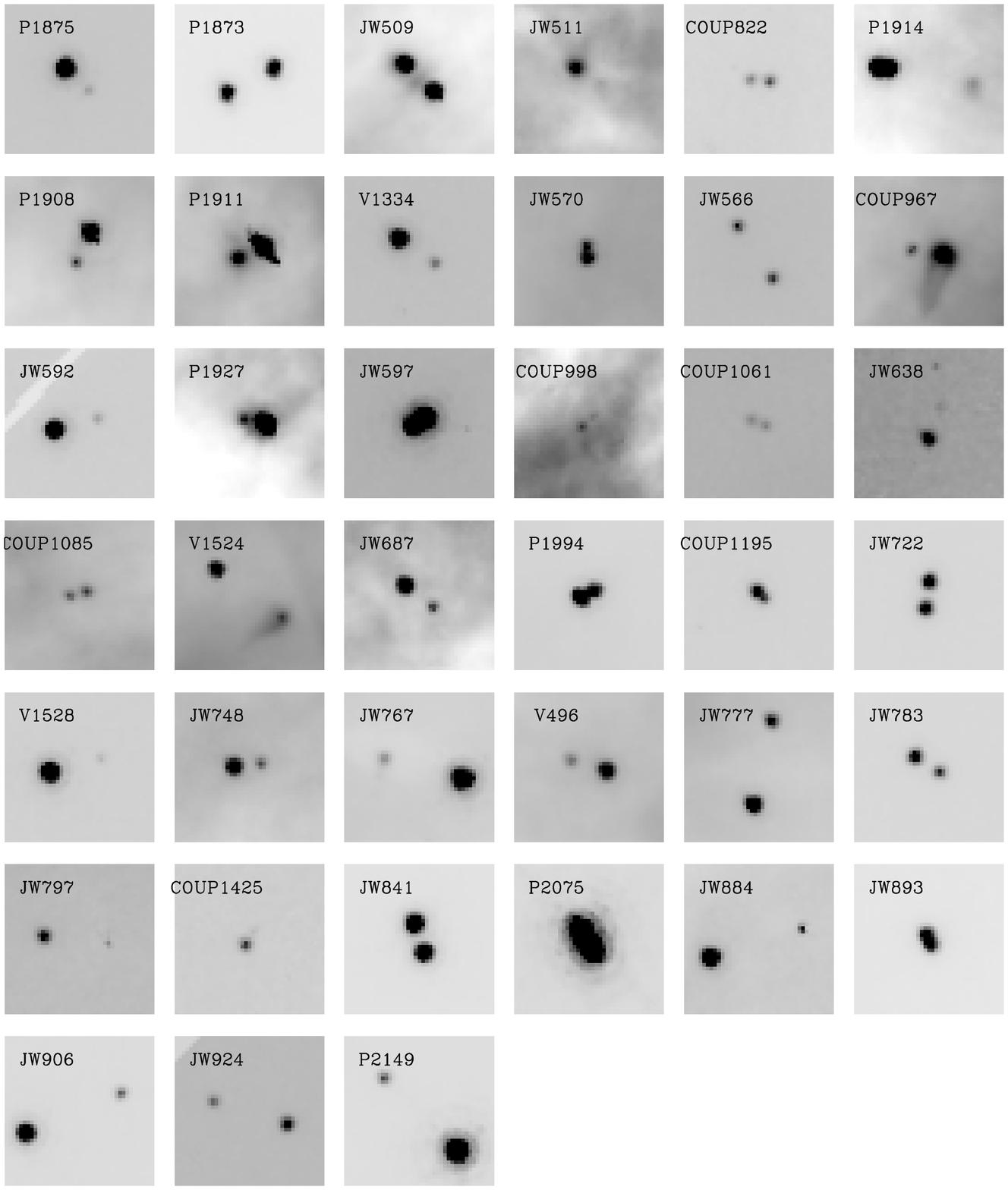}
\caption{continued
\label{fig3b}}
\end{figure}

\clearpage

\begin{figure}[htb]
\epsscale{1.8}
\plotone{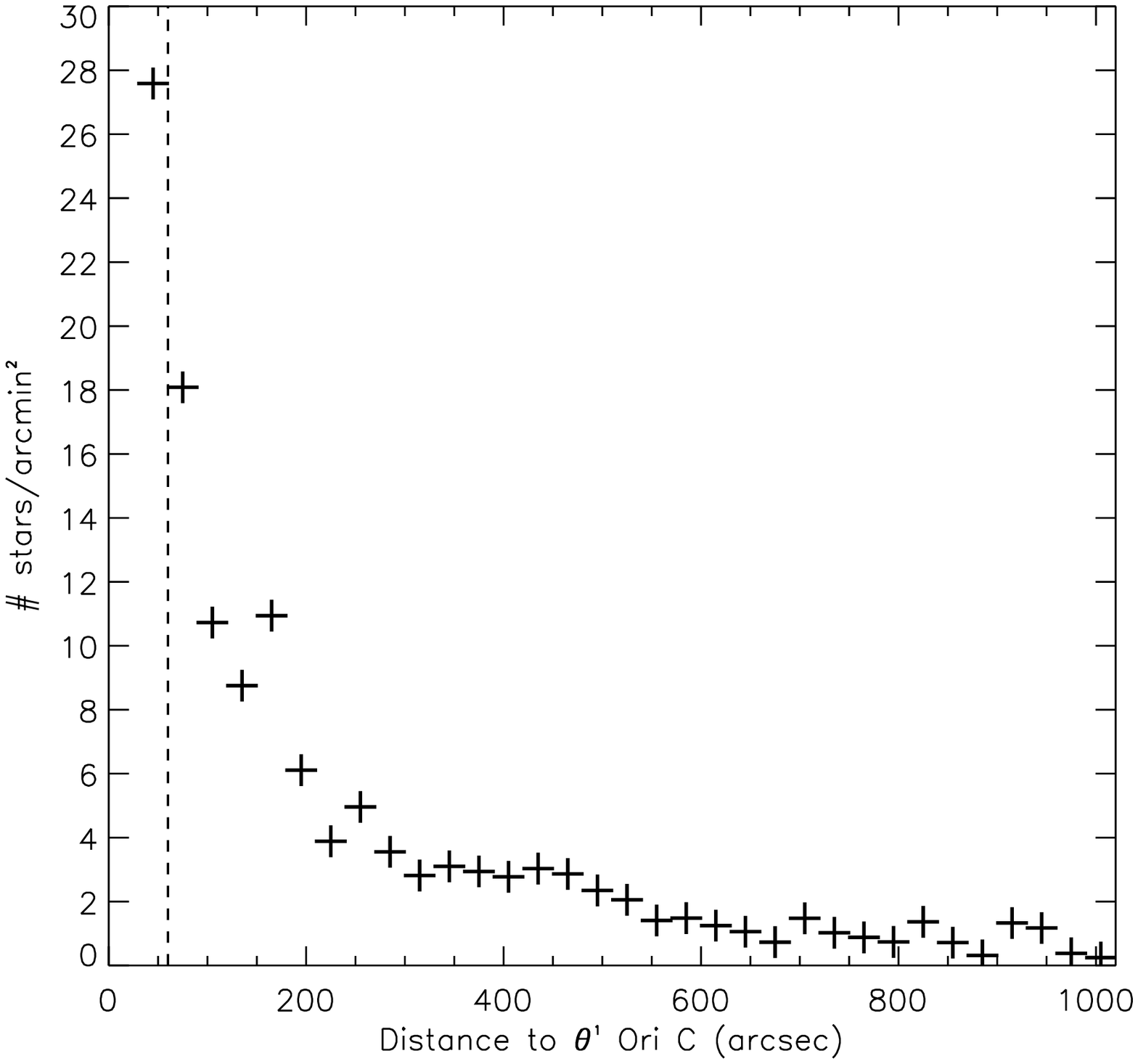}
\caption{The surface density of stars in M42 as a function of distance
from $\theta^1$~Ori~C. Measurements have been done in 30$''$ wide
annuli on our ACS images. The vertical dotted line indicates the
60$''$ exclusion zone inside which we did not attempt to identify
visual binaries due to the high density of stars.
\label{fig4}}
\end{figure}

\clearpage

\begin{figure}[htb]
\epsscale{1.8}
\plotone{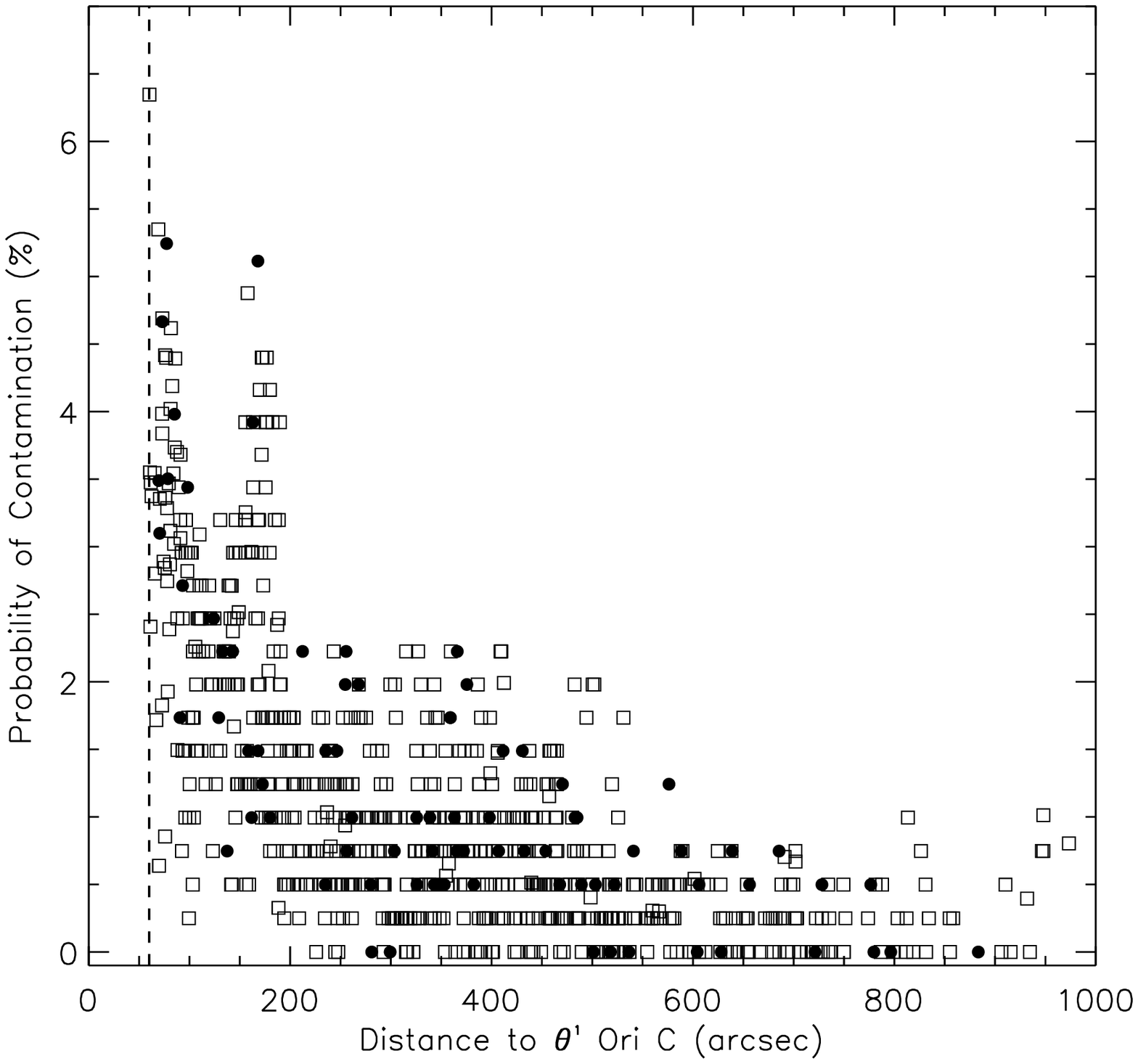}
\caption{The probability that a star would have another star as a
line-of-sight association within a separation of 1$\farcs$5 for all
781 ONC members in our list and as a function of distance from
$\theta^1$~Ori~C. White squares are single stars and black circles are
confirmed binaries. The peak around 180$''$ is due to a
subclustering of stars. 
\label{fig5}}
\end{figure}

\clearpage

\begin{figure}[htb]
\epsscale{1.8}
\plotone{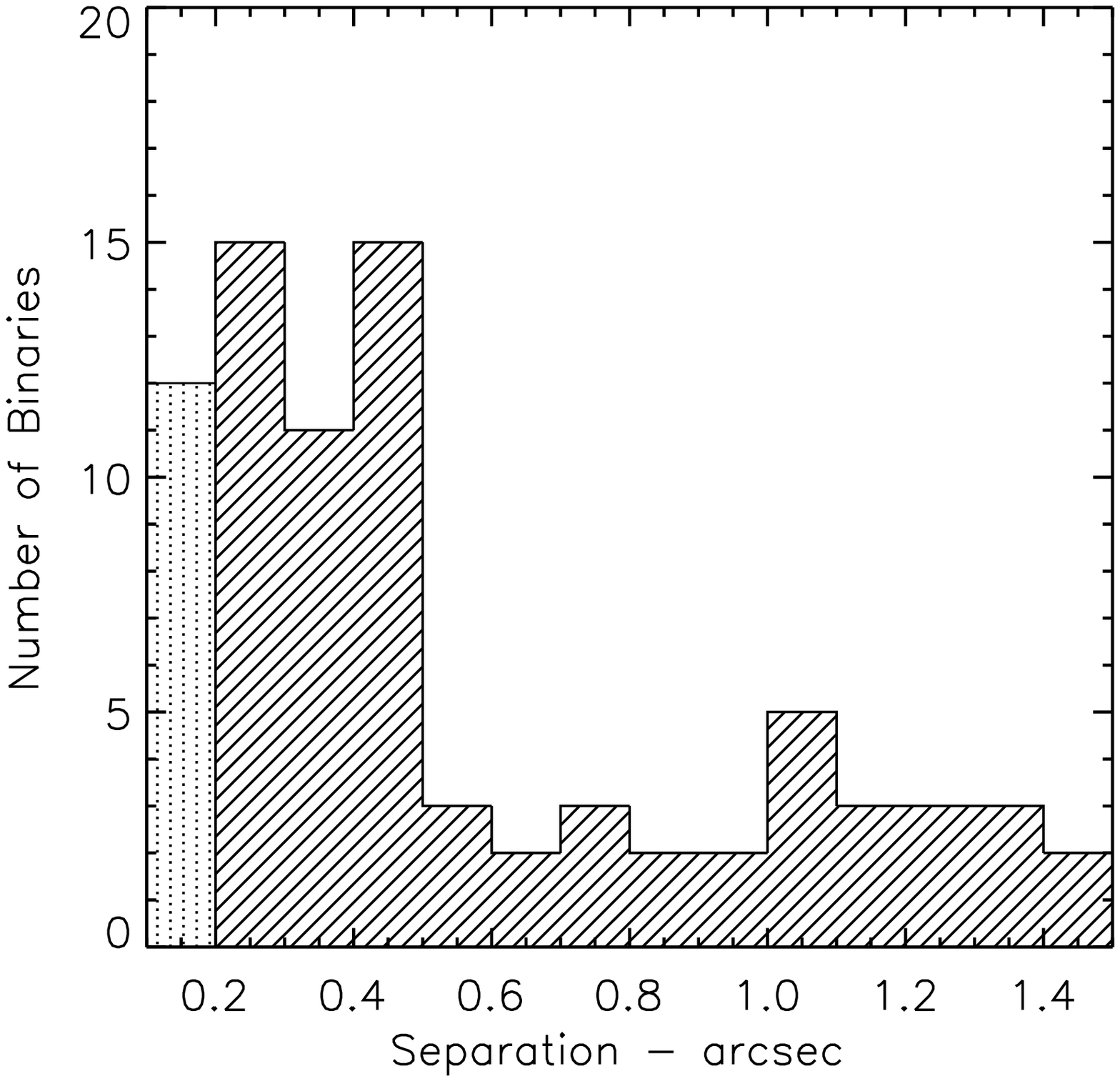}
\caption{ Histogram of binary separations as function of angular
  separation in steps of 0$\farcs$1. The number of binaries in the
  innermost bin with separations less than 0$\farcs$1 is
  incomplete. There is a dramatic decrease in the number of binaries
  when the separation increases beyond 0$\farcs$5.
\label{fig6}}
\end{figure}

\clearpage

\begin{figure}[htb]
\epsscale{1.8}
\plotone{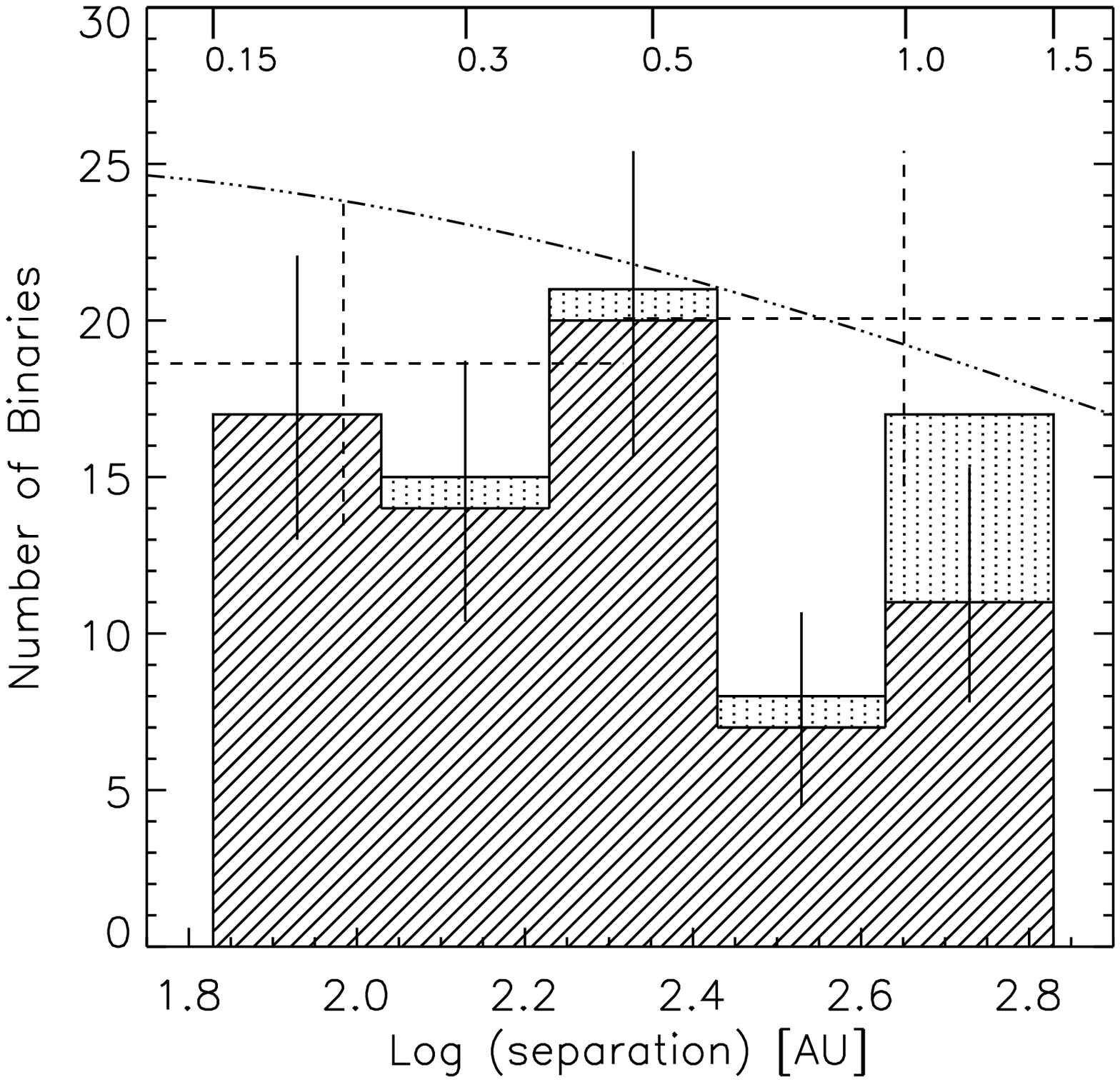}
\caption{ Logarithmic separation distribution function of ONC binaries
  compared to the distribution of field binaries from Duquennoy \&
  Mayor (1991) across the separation range from 0$\farcs$15 to
  1$\farcs$5.  The actual data from Duquennoy \& Mayor within our
  observed range are marked by two dashed crosses and the Gaussian
  distribution that they fit to their complete data set is indicated
  by the dot-dashed curve.  The parts of the columns that are dotted
  indicate the 9 binaries that we calculate are due to line-of-sight
  associations.  The axis on top indicates separations in arcseconds
  at the distance of the ONC.  A scaling was done to correct for the
  fact that Duquennoy \& Mayor has a smaller sample size (164 vs. 781)
  and a larger bin width (0.667 vs. 0.2 in log (AU)), hence a factor
  of (781/164)$\times$(0.2/0.667) = 1.43 was applied to compare the
  distributions directly.
\label{fig7}}
\end{figure}

\clearpage

\begin{figure}
\epsscale{1.8}
\plotone{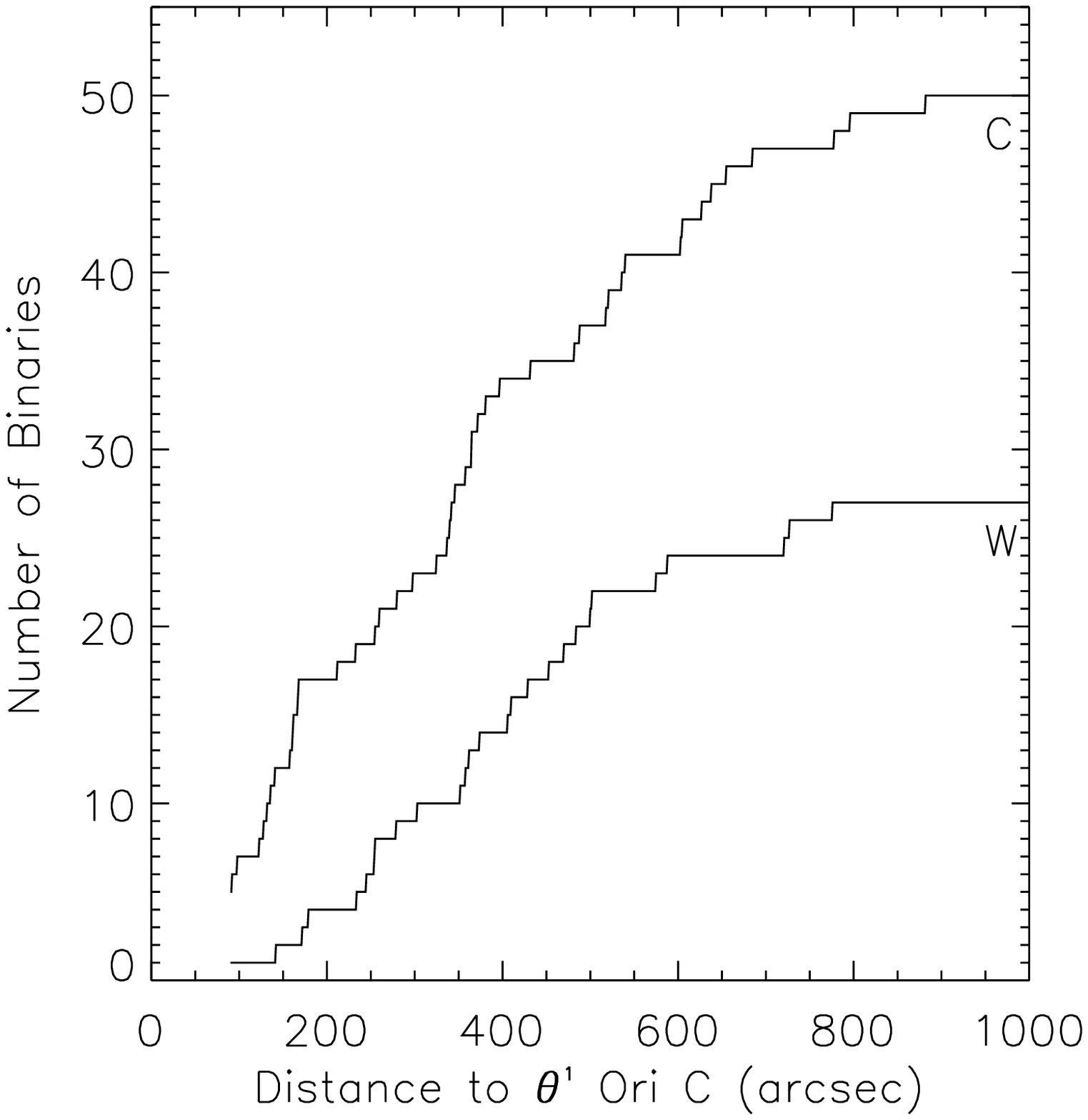}
\caption{Cumulative distributions of close (0$\farcs$15 - 0$\farcs$5) and
  wide (0$\farcs$5 - 1$\farcs$5) binaries in the ONC as a function of
  distance from $\theta^1$~Ori~C.
\label{fig8}}
\end{figure}

\clearpage

\begin{figure}
\epsscale{1.8}
\plotone{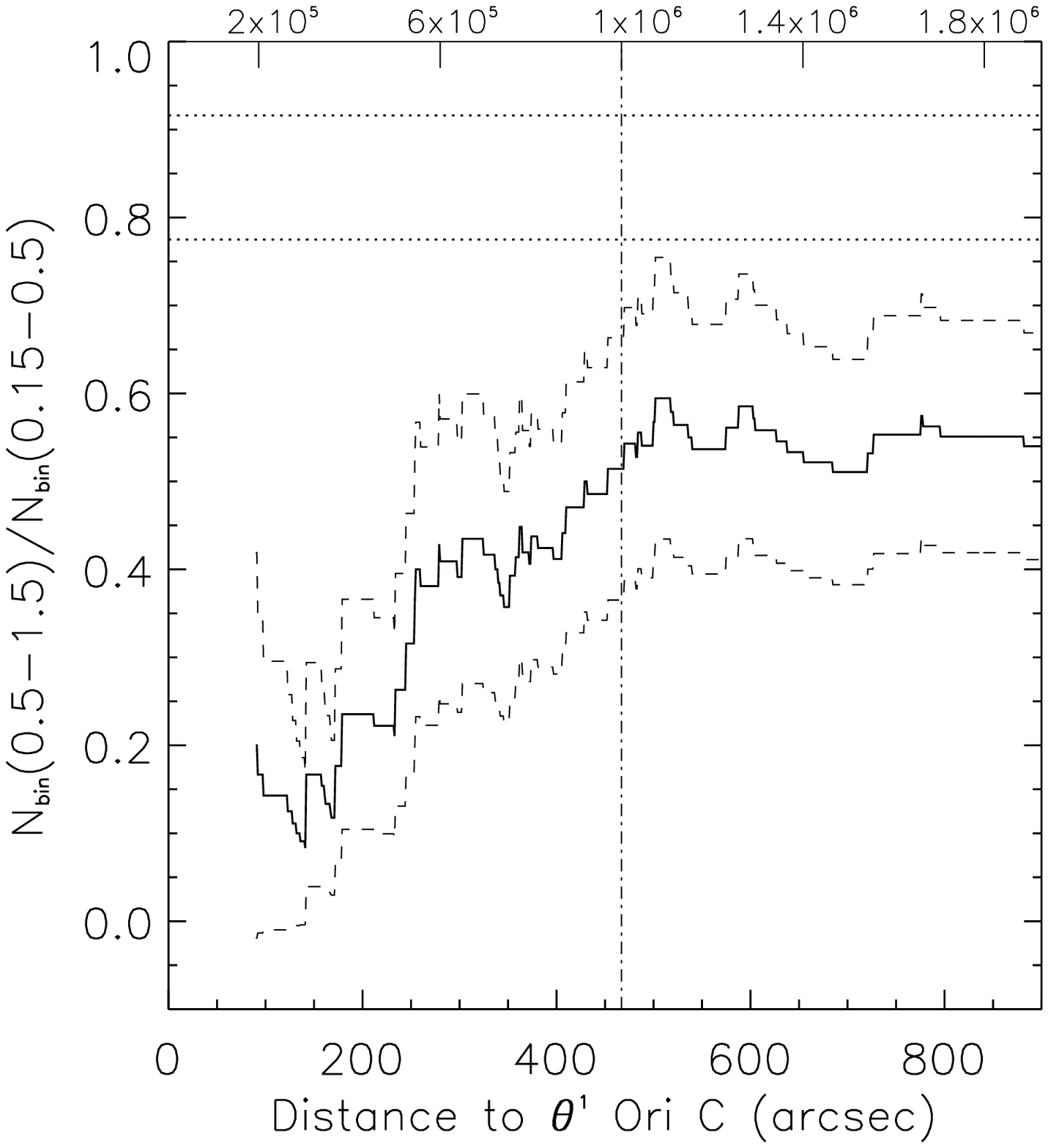}
\caption{The ratio of wide (0$\farcs$5 - 1$\farcs$5) to close (0$\farcs$15 -
  0$\farcs$5) binaries in the ONC as a function of distance to
  $\theta^1$~Ori~C. The figure shows this ratio for all binaries from
  the 60$''$ exclusion zone and out to a given distance. The last
  value at around 900$''$ thus represents the value of {\em all}
  binaries within the above ranges in the ONC outside the exclusion
  zone. The dashed lines indicate the errors on the numbers. The
  dotted horizontal lines represent the same ratio from the Duquennoy
  \& Mayor (1991) field binary study, with the lower being from the
  Gaussian fit, and the upper from their actual data points.  The
  upper scale indicates the crossing time in years, assuming a mean
  one-dimensional velocity dispersion of 2 km/sec. The vertical line
  indicates the distance where the ratio becomes flat, suggesting an
  age for the ONC of about 10$^6$ yr.
\label{fig9}}
\end{figure}

\clearpage

\begin{figure}
\epsscale{1.8}
\plotone{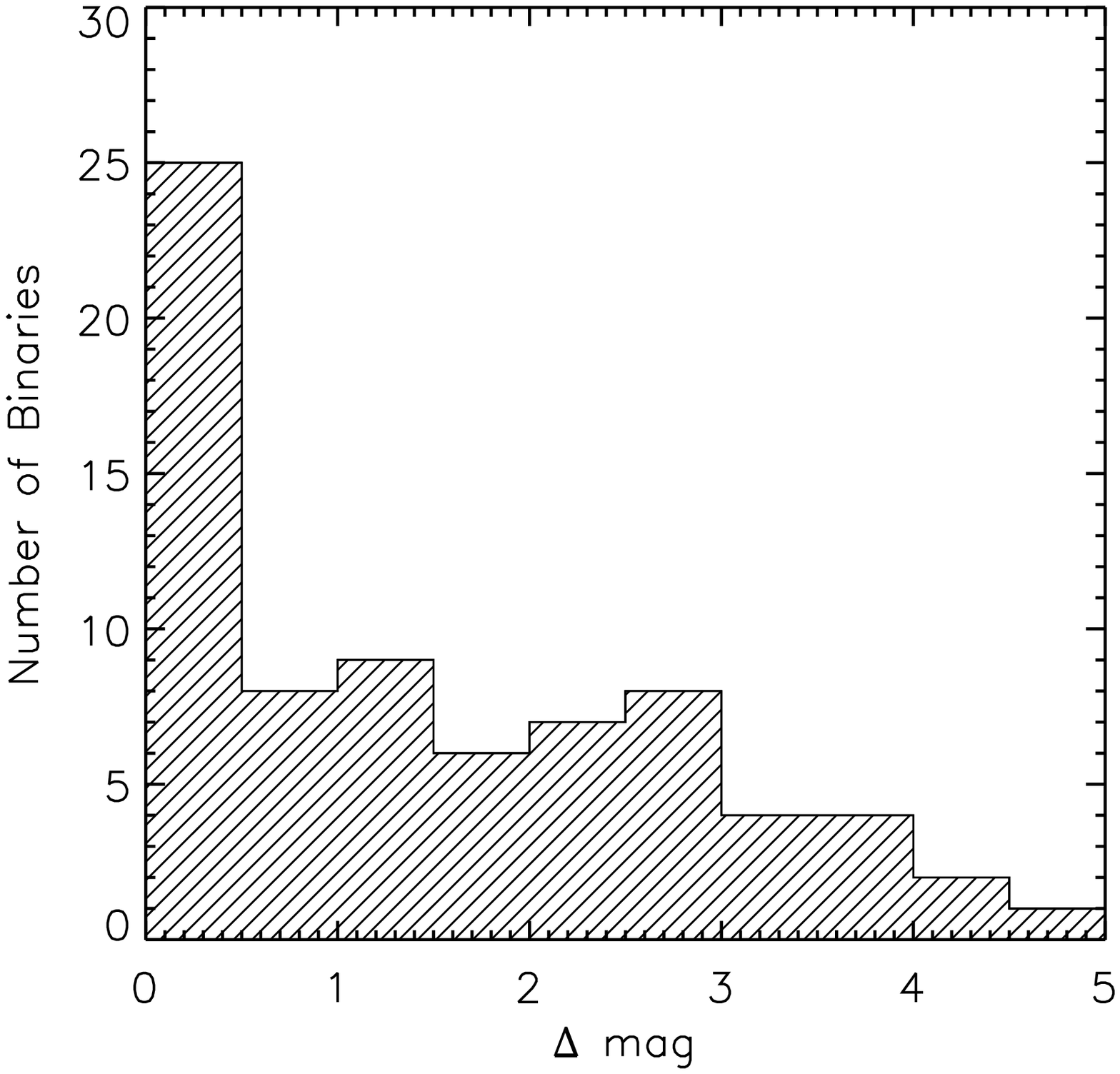}
\caption{Luminosity ratio of the binary population, based on the
  H$\alpha$ fluxes of primaries and secondaries. All stars with
  separations between 0$\farcs$1 and 1$\farcs$5 from Tables~1 and 2
  are plotted, except if they are saturated.
\label{fig10}}
\end{figure}

\end{document}